\documentclass[12pt,a4paper]{article}
\usepackage[T1]{fontenc}
\usepackage{lmodern}
\usepackage[english]{babel}
\usepackage[margin=2.5cm]{geometry}
\usepackage{amsmath,amssymb,graphicx,booktabs,array,longtable,multirow}
\usepackage{float,algorithm,algpseudocode,caption}
\usepackage{placeins}
\usepackage{xcolor,url}
\usepackage[authoryear,round,sort&compress]{natbib}
\usepackage[hidelinks]{hyperref}
\graphicspath{{figures/}{supp_figures/}}

\newcommand{\emp}{\mathrm{emp}}
\DeclareRobustCommand{\src}[1]{\nolinkurl{#1}}

\expandafter\def\expandafter\UrlBreaks\expandafter{\UrlBreaks\do\_\do\-\do\.\do\/}
\allowdisplaybreaks[2]
\hypersetup{pdftitle={ACOR: incremental evidence accumulation towards real-time DNA storage reconstruction},
  pdfauthor={Linxuan Han, Huaming Wu, Shi Jia and Guanjin Qu}}
\newcommand{\ACORStartSupplement}{%
  \clearpage\appendix
  \setcounter{section}{0}\setcounter{subsection}{0}\setcounter{subsubsection}{0}%
  \setcounter{figure}{0}\setcounter{table}{0}\setcounter{equation}{0}%
  \renewcommand{\thesection}{S\arabic{section}}%
  \renewcommand{\thesubsection}{\thesection.\arabic{subsection}}%
  \renewcommand{\thefigure}{S\arabic{figure}}%
  \renewcommand{\thetable}{S\arabic{table}}%
  \renewcommand{\theequation}{S\arabic{equation}}%
  \renewcommand{\theHsection}{supp.\arabic{section}}%
  \renewcommand{\theHsubsection}{supp.\arabic{section}.\arabic{subsection}}%
  \renewcommand{\theHsubsubsection}{supp.\arabic{section}.\arabic{subsection}.\arabic{subsubsection}}%
  \renewcommand{\theHfigure}{supp.\arabic{figure}}%
  \renewcommand{\theHtable}{supp.\arabic{table}}%
  \renewcommand{\theHequation}{supp.\arabic{equation}}%
  \graphicspath{{supp_figures/}}}

\title{ACOR: incremental evidence accumulation towards real-time DNA storage reconstruction}
\author{%
\small Linxuan Han$^{1}$, Huaming Wu$^{1,2}$, Shi Jia$^{3}$ and Guanjin Qu$^{1}$\\
\footnotesize $^{1}$Center for Applied Mathematics, KL-AAGDM, Tianjin University, China\\
\footnotesize $^{2}$State Key Laboratory of Synthetic Biology, Tianjin University, China\\
\footnotesize $^{3}$School of Electronic Information Engineering, Tianjin University, China}
\date{}

\begin{document}
\maketitle
\begin{abstract}
\noindent
\textbf{Motivation:} DNA offers high-density, durable storage by encoding digital information into synthetic strands and retrieving it through sequencing. Sequence reconstruction combines noisy copies of each strand before downstream decoding. Conventional offline methods typically reconstruct completed read clusters, separating reconstruction from ongoing read collection. Streaming reconstruction offers a way to overlap evidence processing with sequencing and reduce the computation remaining before downstream decoding.
\par\smallskip\noindent

\noindent\textbf{Results:} We present ACOR, a streaming consensus reconstruction method that records $k$-mer transitions and boundary support in a persistent ledger, then decodes through length-constrained beam search and priority-based selection at cluster closure. Evaluation on real and simulated DNA-storage datasets showed competitive recovery. On our own short-read next-generation sequencing (NGS) wet-lab library, ACOR combined near-complete recovery with second-scale reconstruction of more than ten million reads. A smaller public benchmark was completed in a fraction of a second. Parallel ACOR achieved speedups of tens to over a thousand-fold relative to the tested serial baseline configurations. Sparse storage reduced high-context memory requirements, while parallel execution accelerated reconstruction computation. ACOR is designed for streaming readout pipelines that process assigned reads while sequencing continues.

 \noindent \textbf{Availability and Implementation:}
ACOR is implemented in C++17/Python for Linux and is available at \url{https://github.com/Jarvis5272/ACOR-Real-Time/releases/tag/v0.2.0}.
\par\smallskip\noindent


 \noindent\textbf{Supplementary information:} Supplementary information accompanies this manuscript.
\end{abstract}
\par\medskip
\section{Introduction}
\label{sec:introduction}

Synthetic DNA combines high information density with long-term stability and is therefore an attractive medium for archival storage~\citep{qu2025dna}. Digital information is encoded into designed DNA strands and recovered by sequencing many molecular copies~\citep{yan2025dna}. Synthesis, amplification and sequencing introduce substitutions, insertions and deletions, so reliable readout depends on accurate recovery of the designed strands~\citep{erlich2017dnafountain,lopez2019dnaassembly,li2023multiple}.

In a DNA-storage readout pipeline, upstream indexing or clustering seeks to group noisy reads by their designed strand, providing per-strand observations for reconstruction \citep{qu2022clover,antkowiak2020lowcost,gimpel2026codecbenchmark}. Reconstruction methods then combine the imperfect observations within each assigned cluster into a consensus sequence, accounting for substitutions and synchronization errors before downstream decoding.

Reconstruction approaches balance recovery quality and computational cost. Bitwise-majority alignment with lookahead (BMALA) corrects local synchronization errors, whereas Viswanathan--Swaminathan (VS) reconstruction combines segment-wise voting with local realignment \citep{gopalan2018trace,viswanathan2008improved}. MUSCLE supplies probability-based multiple sequence alignment for consensus extraction \citep{edgar2022muscle5}. Partial-order-alignment pipelines provide graph-based alternatives: SPOA uses single-instruction, multiple-data (SIMD) acceleration, abPOA adds adaptive banding to reduce alignment work, and POASTA uses A* search for exact gap-affine alignment, followed here by column-wise consensus extraction \citep{vaser2017racon,gao2021abpoa,vandijk2025poasta}. Bidirectional beam search (BBS) estimates cluster-specific Markov probabilities, avoiding alignment and training while supporting efficient recovery \citep{gu2025bbs}. The classical conditional-probability-logic (CPL) component of Deep-DNA estimates conditional probabilities from edit-operation evidence and searches a weighted graph for a consensus without a trained neural model \citep{barlev2025scalable}. ConCluD integrates read partitioning with de Bruijn graph traversal to generate consensus sequences at scale \citep{demoor2025conclud}. Processing arriving reads additionally requires incremental evidence handling.

In nanopore sequencing, basecalled reads can become available during a run \citep{lopez2019dnaassembly}. This enables streaming reconstruction to overlap evidence processing with read acquisition. However, streaming reconstruction faces two challenges: recomputing a consensus after each arrival repeatedly revisits earlier reads, whereas committing a sequence too early can exclude alternative paths supported by later evidence. The central question is therefore how to accumulate evidence as reads arrive while retaining alternatives until a cluster closes.

We present ACOR, which updates a persistent ledger of $k$-mer transitions, boundary support and length statistics as reads arrive, then combines length-constrained beam search with a fixed candidate-selection hierarchy to reconstruct one sequence at cluster closure. It is designed for streaming readout pipelines in which sequencing and upstream clustering incrementally supply strand-assigned reads. The main contributions are:
\begin{itemize}
\setlength{\itemsep}{2pt}\setlength{\parsep}{0pt}
\item \textbf{Bringing reconstruction into ongoing read acquisition.} Incremental $k$-mer evidence accumulation allows reconstruction work to overlap read collection without rescanning earlier reads. This reduces the computation remaining before downstream decoding while retaining later evidence until consensus selection at cluster closure.
\item \textbf{Making high-context, large-scale reconstruction practical.} Sparse storage reserves counters only for observed contexts and transitions, removing the exponential allocation barrier of dense $k$-mer counters. By integrating parallel cluster processing with evidence accumulation and decoding, ACOR achieves near-linear compute acceleration when parallel execution overhead is kept acceptable.
\item \textbf{Near-best recovery with substantially faster reconstruction.} On our NGS wet-lab library, ACOR achieved recovery close to the best tested reconstruction methods. It processed over ten million reads within seconds, with recorded speedups from tens to over a thousand-fold versus tested serial configurations.
\end{itemize}

 \par \section{Methods}
\label{sec:methods}

\subsection{Algorithm overview}
\label{sec:overview}

ACOR reconstructs a non-empty cluster of reads assigned to one designed strand, using a design length $L$ supplied by the upstream readout protocol. Read assignment and the cluster-closure signal are external inputs; the latter marks the end of that cluster's read input. Each normalized read updates the evidence ledger; decoding occurs once, at cluster closure.

\begin{figure}[ht!]
\centering

\includegraphics[width=\linewidth]{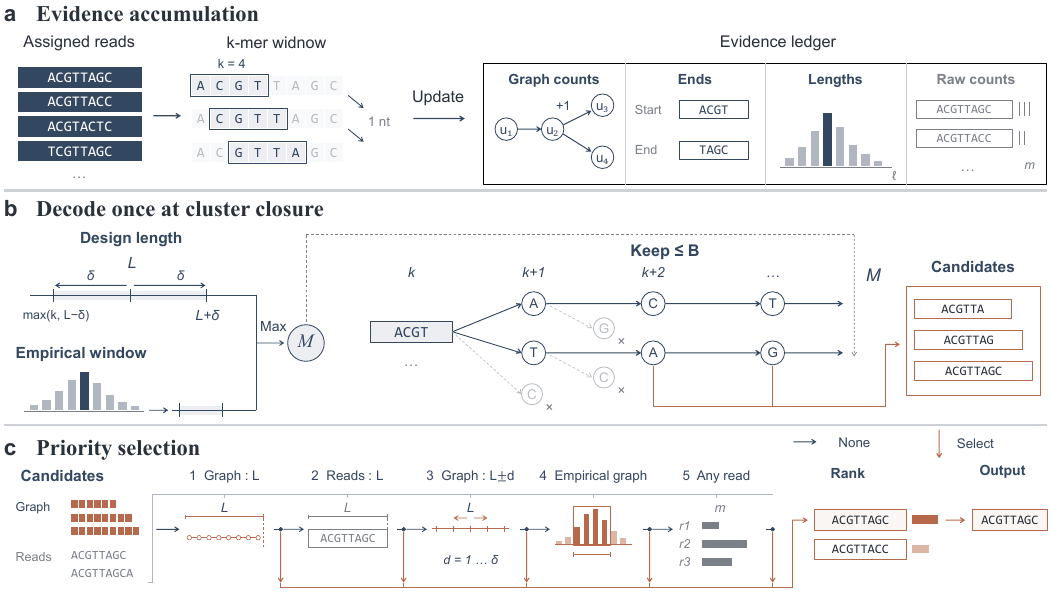}

\caption{ACOR reconstruction workflow. Incoming reads update a cluster-level evidence ledger. After cluster closure, admissible lengths and supported starting contexts are determined, a bounded search traverses observed transitions and caches candidates at admissible lengths, and a fixed priority hierarchy selects one output sequence. Each design-length cache holds at most $B$ candidates; empirical lengths share one cache. Candidates are cached before extension at each depth. Neighbouring lengths are jointly ranked at increasing offsets; selection returns from the first non-empty family. Grey paths denote ultimately discarded alternatives.}
\label{fig:overview}
\end{figure}

ACOR combines incremental evidence accumulation with sequence-graph and bounded-search ideas in three stages (Fig.~\ref{fig:overview}) \citep{idury1995assembly,pevzner2001eulerian,malde2005wordgraph,gu2025bbs,demoor2025conclud,muthukrishnan2005streams}. First, a length-$k$ sliding window extracts contexts and single-base transitions from each read to update the ledger. Second, after the current read cluster closes, the ledger defines admissible lengths and supported roots. A bounded search follows observed transitions, retains at most $B$ active prefixes per depth, and caches candidates at admissible lengths. Third, a fixed hierarchy returns one sequence from the first non-empty candidate family.

Here, $N$ denotes cluster read count, $k$ context length and $B$ beam width. We choose $k\leq L$ to fit a complete $k$-mer within the target strand. Parameter values and execution details appear under Comparison settings and in Supplementary Note~\ref{sec:implementation}.

 \subsection{Incremental accumulation of sequence evidence}

A designed DNA strand can be represented as a string over the nucleotide alphabet $\Sigma=\{A, C, G, T\}$. ACOR processes its non-empty noisy reads one at a time. Before extracting evidence, each read is normalized by converting valid nucleotide symbols to the canonical alphabet and mapping all other symbols to $\mathtt N$. Write the normalized read as $x=x_1x_2\cdots x_{|x|}$.

To obtain $k$-mers, ACOR slides a length-$k$ window over $x$ one position at a time. Each valid window is treated as a context. If a following canonical base exists, the next window differs by deleting the first base and appending that base; this defines an observed transition:
\begin{equation}
u_i=x_i x_{i+1}\cdots x_{i+k-1},
\qquad
u_i\xrightarrow{\;x_{i+k}\;}
u_{i+1}=x_{i+1}\cdots x_{i+k},
\label{eq:window}
\end{equation}
where $i=1,\ldots,|x|-k+1$ for contexts and $i=1,\ldots,|x|-k$ for transitions. Sweeping the window across all reads therefore defines an implicit sequence graph $\mathcal G=(V,E)$: vertices are observed contexts, directed edges are observed single-base extensions, and edge multiplicities record their support. The graph is represented by these accumulated counts rather than by repeatedly materializing a separate graph for every read.

Instead of decoding after each update, ACOR maintains a cluster-local evidence ledger. After $t$ reads, the ledger is
\begin{equation}
Z^{(t)}=
\bigl(v_t,n_t,o_t,a_t,h_t,\ell_t,m_t\bigr),
\label{eq:ledger}
\end{equation}
where the components have four roles. The graph evidence consists of context counts $v_t(u)$ and transition counts $n_t(u,b)$. The cached outgoing total
\[
o_t(u)=\sum_{b\in\Sigma}n_t(u,b)
\]
supports later transition scoring without repeated summation. Boundary evidence is recorded by $a_t(u)$ and $h_t(u)$, which count reads whose first and last valid contexts are $u$, respectively. The length summary $\ell_t$ stores the exact median read length together with running sums of log lengths and squared log lengths. Finally, $m_t(r)$ records the frequency of each normalized raw read $r$.

For the $t$-th normalized read $r_t$, let $V(r_t,u)$ and $E(r_t,u,b)$ denote the occurrences of context $u$ and transition $(u,b)$ within that read. The graph counts are updated by
\begin{align}
v_t(u)&=v_{t-1}(u)+V(r_t,u),\\
n_t(u,b)&=n_{t-1}(u,b)+E(r_t,u,b).
\label{eq:update}
\end{align}
If $\operatorname{first}(r_t)$ and $\operatorname{last}(r_t)$ denote the first and last valid contexts in $r_t$, then
\begin{align}
a_t(u)&=a_{t-1}(u)+\mathbf{1}\{u=\operatorname{first}(r_t)\},\\
h_t(u)&=h_{t-1}(u)+\mathbf{1}\{u=\operatorname{last}(r_t)\},\\
m_t(r)&=m_{t-1}(r)+\mathbf{1}\{r_t=r\}.
\end{align}
The length component is updated from the current read length:
\[
\ell_t=
\left(
\operatorname{med}_{j\leq t}|r_j|,
\sum_{j=1}^{t}\log |r_j|,
\sum_{j=1}^{t}(\log |r_j|)^2
\right).
\]

The median is maintained with exact length samples or two heaps; the displayed tuple summarizes this state rather than its full storage representation (Supplementary Note~S1).

A read containing no valid context contributes only to its length statistics and normalized-string frequency. Thus, every read updates the ledger once, while no partial consensus is produced and no earlier read is rescanned.

\subsection{Evidence-guided beam search}
\label{sec:decode}
\label{sec:decoder}

After incorporating the closing read, ACOR uses the completed cluster ledger to set admissible lengths, choose starting contexts and grow candidate sequences. Design length and observed read lengths provide the two sources of length constraints.

The upstream protocol supplies design length $L$, whereas context length $k$ is preset. The constants $\sigma_0,\tau,\rho_\pm$ and $\delta$ remain fixed within each configuration (Supplementary Note~\ref{sec:implementation}). Cluster read lengths determine $\widetilde L$ (median length), $\mu$ (mean log length) and $\sigma$ (population standard deviation of log lengths, floored at $\sigma_0>0$). The empirical window is
\begin{align}
L_{\emp}&=
\max\left\{
k,\lfloor e^{\mu-\tau\sigma}\rfloor,
\lfloor\rho_-\widetilde L\rfloor
\right\},\\
U_{\emp}&=
\max\left\{
L_{\emp},
\min\left(
\lceil e^{\mu+\tau\sigma}\rceil,
\lceil\rho_+\widetilde L\rceil
\right)
\right\},
\label{eq:length-bounds}
\end{align}
where $\tau>0$ controls the spread of the empirical range and $0<\rho_-\leq\rho_+$ limits its relation to the median read length. The design window is
\[
W_D=[\max(k,L-\delta),L+\delta],
\]
where $\delta$ is a non-negative length tolerance. The maximum searched length is
\[
M=\max(L+\delta,U_{\emp}).
\]
The windows determine which prefix lengths can be cached as candidates; $M$ sets the extension limit. Lengths below $k$ are excluded because they cannot contain a complete context. The windows and $M$ are derived, not independently tuned.

ACOR next ranks contexts by their support as the first valid contexts in reads and retains at most $B$ starting contexts. If start support is unavailable, overall context counts provide the fallback ranking. Let $\widetilde a(u)$ denote the resulting start support and $V_s$ the contexts with positive support. Their total support is $A=\sum_{u\in V_s}\widetilde a(u)$. The initial and transition scores are
\begin{align}
s_0(u)&=
\log\frac{\widetilde a(u)+1}{A+|V_s|}
+\alpha\log(1+\widetilde a(u)),\\
q(u,b)&=
\log\frac{n(u,b)+1}{o(u)+4}
+\alpha\log(1+n(u,b)),
\label{eq:local-scores}
\end{align}
Each score combines a log-probability term derived from add-one-smoothed support counts with an $\alpha$-weighted reward for repeated observations. Each retained root initializes a length-$k$ prefix, which grows one base at a time along observed transitions with $n(u,b)>0$. The next active beam retains at most $B$ prefixes ranked by cumulative score $P$. For a path ending in context $e(p)$,
\begin{align}
P(pb)&=P(p)+q(e(p),b),\\
S(p)&=
\frac{P(p)}{|p|-k+1}
-\lambda_{\mathrm{len}}
\left(\frac{\log |p|-\mu}{\sigma}\right)^2
+\lambda_{\mathrm{end}}\log(1+h(e(p))),
\label{eq:path-scores}
\end{align}
Here, $P(u)=s_0(u)$ initializes each root score. $P$ controls which prefixes continue, whereas $S$ ranks complete candidates at admissible output lengths. $S$ combines normalized path support, length plausibility and terminal evidence. The weights $\lambda_{\mathrm{len}}$ and $\lambda_{\mathrm{end}}$ control the length penalty and terminal-support reward in $S$, not in beam pruning.

At each depth, admissible-length prefixes are cached before any further extension. Design-window candidates are cached separately by length, whereas all empirical-window lengths share one cache. Each cache retains at most $B$ distinct strings ranked by $S$. The active beam independently retains at most $B$ prefixes ranked by $P$. Pruning a search branch does not itself remove a previously cached candidate. The search ends at $M$ or when no active prefix remains, then passes its caches to consensus selection. Score rounding and lexical tie-breaking follow the deterministic ordering specified in Supplementary Note~S1.

\subsection{Priority-based consensus selection}
\label{sec:selection}

ACOR checks candidate families in a fixed order and returns one consensus sequence (Fig.~\ref{fig:overview}c). Let $\mathcal C_\ell$ denote the design-window cache at length $\ell$ and $\mathcal C_{\emp}$ the shared empirical cache. Missing caches and lengths below $k$ are treated as empty. Let $\mathcal R=\{r:m(r)>0\}$ be the normalized read strings and $\mathcal R_\ell=\{r\in\mathcal R:|r|=\ell\}$. For integer $\delta\geq0$, the family order is
\begin{equation}
\begin{aligned}
\mathcal Q_1&=\mathcal C_L,\qquad
\mathcal Q_2=\mathcal R_L,\\
\mathcal Q_{d+2}&=\mathcal C_{L-d}\cup\mathcal C_{L+d},
\quad d=1,\ldots,\delta,\\
\mathcal Q_{\delta+3}&=\mathcal C_{\emp},\qquad
\mathcal Q_{\delta+4}=\mathcal R.
\end{aligned}
\label{eq:candidate-families}
\end{equation}
ACOR examines these families in order, skips empty ones and returns the top-ranked sequence from the first non-empty family. Later families are not compared once this selection is made. At each increasing offset $d$, candidates at $L-d$ and $L+d$ are ranked together; neither side is preferred in advance. With $\delta=0$, the neighbouring-length families are omitted. Graph candidates are ranked by $S$, using the score ordering in Supplementary Note~S1. Normalized read strings are ranked by decreasing integer frequency $m$ and then lexical order. The final raw-read family is not limited by the length windows and supplies an output for a successfully processed non-empty cluster.

Algorithm~\ref{alg:acor} summarizes the procedure. Here, $F$ is the active prefix set, $\mathcal H$ its one-step extensions and $\mathcal C$ the candidate caches.

\begin{algorithm}[H]
\caption{ACOR reconstruction of one cluster}
\label{alg:acor}
\small
\begin{algorithmic}[1]
\Require Stream of $(x,\mathrm{close})$ events; $L\geq k$; beam width $B$; fixed scoring settings
\State $Z\gets\Call{EmptyLedger}{}$
\Repeat
  \State $(x,\mathrm{close})\gets\Call{Next}{\mathrm{stream}}$
  \State $r\gets\Call{Normalize}{x}$; $Z\gets\Call{Update}{Z,r}$
\Until{$\mathrm{close}$}
\State $(W_D,W_{\emp},M)\gets\Call{Bounds}{Z,L}$
\State $F\gets\Call{Roots}{Z,B}$; $\mathcal C\gets\Call{EmptyCaches}{}$
\State $P(u)\gets s_0(u)$ for each $u\in F$
\For{$d=k,\ldots,M$}
  \State \textbf{if} $F=\varnothing$ \textbf{then break}
  \State $\mathcal C\gets\Call{Cache}{\mathcal C,F,d,W_D,W_{\emp},S,B}$
  \State \textbf{if} $d=M$ \textbf{then break}
  \State $\mathcal H\gets\{pb:p\in F,\ b\in\Sigma,\ n(e(p),b)>0\}$
  \State $P(pb)\gets P(p)+q(e(p),b)$ for each $pb\in\mathcal H$
  \State $F\gets\Call{Keep}{\mathcal H,P,B}$
\EndFor
\State \Return $\Call{Select}{\mathcal C,Z,L,S}$
\end{algorithmic}
\end{algorithm}

\subsection{Computational complexity}
\label{sec:complexity}

For a cluster with $N$ reads, total read length
\[
R=\sum_j |x_j|,
\]
distinct normalized raw-read characters $R_U$, observed graph contexts $|V|$, supported roots $|V_s|$, observed transitions $|E|$, maximum admissible length $M$ and beam width $B$, the main stages have the following expected time cost, excluding representation-dependent initialization:
\begin{align}
\mathbb{E}[T_{\mathrm{cluster}}]
=
O\!\left(
R+N\log(N+1)+|V_s|\log(|V_s|+1)
\right)
\notag\\+
O\!\left(
BM^2\log(B+1)
\right).
\label{eq:stages}
\end{align}
With $U$ the set of distinct normalized read strings, $R_U=\sum_{r\in U}|r|$ is their total length (Supplementary Note~S2).

The terms describe read accumulation, exact length statistics, root ranking and bounded search with materialized prefixes. Fixed $B$ and $M$ remove explicit $N$-dependence from search, not from accumulation or root ranking.

For freshly constructed per-worker state, a dense direct-indexed ledger requires
\begin{equation}
A_{\mathrm{worker}}^{\mathrm{dense}}
=
O\!\left(
4^{k+1}+N+R_U+BM
\right),
\label{eq:dense-memory}
\end{equation}
Dense preparation additionally requires $O(4^{k+1})$ time. In the sparse representation, only observed contexts and transitions are stored:
\begin{equation}
A_{\mathrm{worker}}^{\mathrm{sparse}}
=
O\!\left(
|V|+|E|+N+R_U+BM
\right).
\label{eq:sparse-memory}
\end{equation}
Sparse storage allocates counters for observed evidence; both representations retain length samples and distinct raw-read strings. Complete bounds and parallel accounting are given in Supplementary Notes~S2 and~S6; implementation details are in Supplementary Note~\ref{sec:implementation}.

\section{Results}
\label{sec:results}

\subsection{Datasets}

We evaluated ACOR on real and simulated DNA-storage reads spanning Illumina and nanopore sequencing. The original benchmark suite comprised approximately 8.8 million reads from 341,867 non-empty clusters \citep{lopez2019dnaassembly,chandak2020overcoming,srinivasavaradhan2021trellisbma,barlev2025scalable,welter2026endtoend}. Dataset dimensions, error rates and overlapping subsets are documented in Table~\ref{tab:benchmark-main}.

Within each fixed cluster, reads were permuted under recorded seeds and replayed sequentially. ACOR consumed every read before decoding, and references were reserved for evaluation in these short-sequence benchmarks. Microsoft CNR additionally includes provider-reported sequence dependencies and possible upstream assignment errors \citep{microsoft2021cnrdataset}.

Our own next-generation sequencing (NGS) wet-lab library comprised 96,055 distinct 100-nt payloads synthesized as 129-nt oligos with fixed flanks. Sequencing generated 66,127,110 read pairs (2$\times$150~nt). Merging and trimming yielded 18,191,190 insert reads; reference-assisted assignment retained 17,701,449 across all designs (Table~\ref{tab:benchmark-main}). References served upstream assignment and evaluation, not reconstruction.

Each design cluster contained 3--567 retained read observations (median, 179; 5th--95th percentiles, 86--302). This is post-assignment read depth, not unique-molecule coverage: no UMI deduplication was performed. Processing protocols and record limitations are detailed in Supplementary Note~S5.

\begin{table}[t]
\caption{Dimensions and read-level error statistics of public benchmarks and our own NGS wet-lab library.}
\label{tab:benchmark-main}

\centering
\fontsize{10}{12}\selectfont
\setlength{\tabcolsep}{3pt}
\renewcommand{\arraystretch}{1.05}

\begin{tabular*}{\textwidth}{@{\extracolsep{\fill}}lrrcc@{}}
\toprule
Dataset
& \shortstack{Non-empty\\clusters}
& Reads
& \shortstack{Design length\\(nt)}
& \shortstack{Read-level error\\(\%)} \\
\midrule

oligo0
& 1,466
& 167,546
& 108
& 13.5596 \\

ncomms19 365 dishes
& 2,038
& 66,804
& 150
& 5.9521 \\

binned nanopore
& 10,000
& 213,740
& 140
& 4.3352 \\

Microsoft CNR
& 9,984
& 269,709
& 110
& 5.7670 \\

Trellis BMA
& 7,373
& 81,103
& 110
& 5.8636 \\

DNAFormer nanopore
& 109,976
& 2,354,584
& 140
& 4.3952 \\

DNAFormer Illumina
& 109,944
& 3,183,840
& 140
& 0.1195 \\

Zenodo 10943282
& 91,086
& 2,436,631
& 150
& 4.2513 \\

\textbf{Own NGS wet-lab library}
& \textbf{96,055}
& \textbf{17,701,449}
& \textbf{100}
& \textbf{0.8105} \\
\midrule
Total
& 437,922
& 26,475,406
& --
& --
\\
\bottomrule
\end{tabular*}
\par\vspace{6pt}
\noindent\begin{minipage}{0.98\textwidth}
\footnotesize
\textit{Note:}
Read-level error is the summed read-to-reference edit distance divided by the summed reference lengths, counting one reference length per read. Values describe evaluated reads, not raw sequencing error. The own-NGS row contains assigned read observations and 100-nt payloads. Totals sum evaluated dataset rows, including the overlapping binned nanopore subset of DNAFormer nanopore. Trellis BMA is simulated; Zenodo 10943282 excludes 680 empty reference clusters.
\end{minipage}

\end{table}

\subsection{Comparison settings}
\label{sec:comparison-settings}

All methods received identical normalized clusters and design lengths. Baselines included bidirectional beam search (BBS) \citep{gu2025bbs}, Viswanathan--Swaminathan reconstruction (VS) \citep{viswanathan2008improved,sabary2024reconstruction}, and bitwise-majority alignment with lookahead (BMALA) \citep{gopalan2018trace,sabary2024reconstruction}. We also evaluated the Deep-DNA consensus component (CPL) \citep{barlev2025scalable} and MUSCLE with consensus extraction \citep{edgar2004muscle,edgar2022muscle5}.

ACOR used $k=9$ and $B=16$ unless otherwise stated; P20 denotes 20 workers. Table~\ref{tab:hardware-main} lists the execution environments. Original benchmark experiments, the separate A800 acceptance batch and subsequent optimized-implementation experiments are identified separately. Baseline resource allocations differed, so Fig.~\ref{fig:runtime-all} compares the tested execution configurations rather than equal-worker efficiencies. Full defaults and original execution settings are retained in Supplementary Note~\ref{sec:implementation}.

Controlled scaling used three timing repeats per worker count. Its 20-million-read input was constructed by fourfold replication of a five-million-read synthetic workload (Supplementary Note~S6). Expanded Oxford Nanopore Technologies (ONT) comparisons used reference-guided windows, but consensus methods received assigned reads rather than reference sequences. Their construction, single-observation ACOR timings, baseline repeats and stopping rules are specified in Supplementary Note~S7.

\begin{table}[t]
\caption{Execution environment.}
\label{tab:hardware-main}
\centering
\fontsize{10}{12}\selectfont
\setlength{\tabcolsep}{3pt}
\renewcommand{\arraystretch}{1.00}
\begin{tabular*}{\textwidth}{@{\extracolsep{\fill}}p{30mm}p{40mm}>{\raggedleft\arraybackslash}p{17mm}>{\raggedleft\arraybackslash}p{19mm}p{42mm}@{}}
\toprule
Machine & Processor & Logical CPUs & RAM (GiB) & Operating system \\
\midrule
LEGACY (main runs) & 2$\times$ Xeon E5-2660 v3 & 40 & 125.77 & Ubuntu; kernel 5.15.0-139-generic \\
A800 (additional runs) & 2$\times$ Xeon Silver 4316 & 80 & 503 & Ubuntu 20.04.6 \\
Windows workstation & Core i7-10750H & 12 & 16 & Windows 11 Home, 25H2 \\
\bottomrule
\end{tabular*}

\par\vspace{2pt}\smallskip
\raggedright\footnotesize Logical CPUs are hardware threads; LEGACY and A800 have 20 and 40 physical cores, respectively. RAM is the total capacity reported by the operating system. The A800 machine also has 2$\times$ NVIDIA A800 80 GB GPUs.
\end{table}

\subsection{Evaluation metrics}
For $J$ evaluated clusters, with reference $s_i$, reconstruction $\widehat s_i$ and Levenshtein distance $d_i=\mathrm{ED}(\widehat s_i,s_i)$,
\begin{equation}
\begin{aligned}
\mathrm{Exact}&=\frac{1}{J}\sum_{i=1}^{J}\mathbf{1}\{\widehat s_i=s_i\},\\
\mathrm{Accuracy}&=1-\frac{\sum_{i=1}^{J}d_i}{\sum_{i=1}^{J}|s_i|},\\
\mathrm{Mean\,ED}&=\frac{1}{J}\sum_{i=1}^{J}d_i.
\end{aligned}
\label{eq:evaluation-metrics}
\end{equation}
Exact and Accuracy are expressed as percentages. Missing-output handling is specified in Supplementary Note~S3.

ACOR engine time covers loading, reconstruction and prediction writing, excluding external preparation and evaluation. Baseline timings exclude evaluation but retain native scopes and worker allocations. Controlled scaling uses separate compute time, excluding input/output but retaining initialization and dispatch (Supplementary Notes~S5--S6; Supplementary Note~\ref{sec:implementation}). 

\subsection{Reconstruction fidelity across sequencing datasets}

ACOR achieved competitive exact-sequence recovery across eight public benchmarks spanning experimental and simulated sequencing data (Fig.~\ref{fig:main-results}c). Exact ranged from 92.16\% to 99.92\%, and four datasets were within 0.14 percentage points of the best tested baseline. On the highest-error benchmark, oligo0, ACOR recovered the largest fraction of exact sequences despite 13.56\% read-level error. This advantage was metric-dependent: CPL had higher Accuracy and lower Mean ED on oligo0. On simulated Trellis BMA, ACOR's Exact was 2.79 percentage points below the best baseline (Supplementary Table~S1).

The full-library comparison on our own NGS data reinforced the public-benchmark results (Table~\ref{tab:own-ngs-p16}). ACOR returned one sequence for every one of the 96,055 designs, providing complete output coverage. At $k=15$, it achieved 99.9844\% Exact, and all predictions matched BBS across the entire design pool. Only 15 outputs differed from their references, placing ACOR close to the best tested recovery rate; CPL's Exact was slightly higher.

Increasing $k$ from 9 to 15 reduced non-exact outputs from 75 to 15 and maximum edit distance from 50 to 2. The higher-context configuration therefore improved exact recovery while limiting the largest observed residual error to two edits. These measurements distinguish the quality of the reconstructed sequences from simply returning an output for every design.

\subsection{Fast reconstruction of complete libraries}

ACOR's short recorded runtimes extended from public benchmarks to the complete NGS wet-lab library. Its $k=9$/P40 configuration processed 17,701,449 assigned read observations and returned one sequence per design in 9.01~s, with 99.9219\% Exact. This parallel configuration achieved approximately 28--1,330-fold acceleration relative to seven tested serial baseline configurations. The higher-recovery configuration completed the library in 20.53~s in the P16 comparison reported in Table~\ref{tab:own-ngs-p16}, which pairs quality, timing and memory for each configuration.

\begin{figure}[t]
\centering

\includegraphics[width=\textwidth]{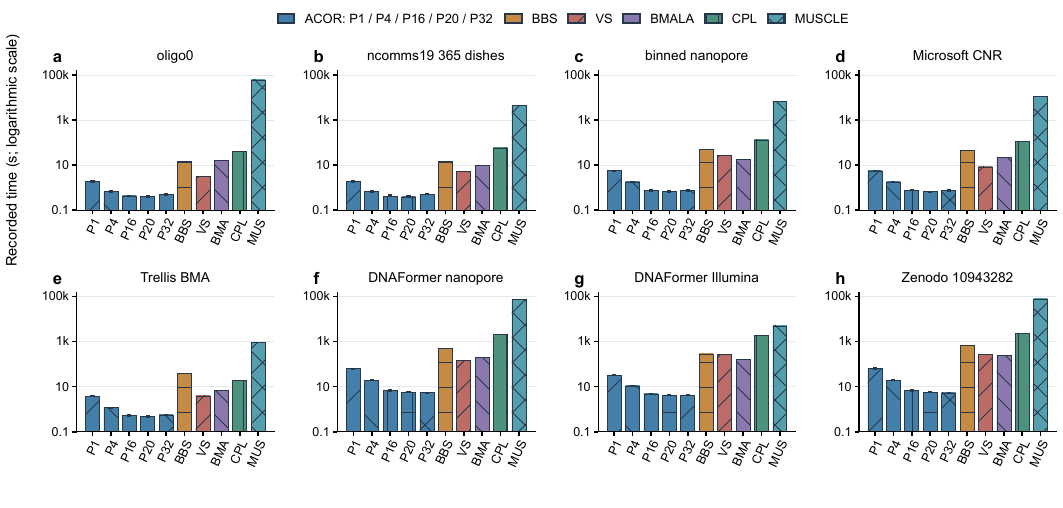}

\caption{Reconstruction time on experimental and simulated datasets. Axes are logarithmic. ACOR bars show three-seed median engine times, with whiskers spanning the observed minima and maxima. Baseline bars retain the original method-specific timing estimates; MUSCLE was measured once. All times exclude evaluation, and methods use different native resource allocations. P denotes ACOR workers; BMA and MUS abbreviate BMALA and MUSCLE.}
\label{fig:runtime-all}
\end{figure}

\begin{figure}[!t]
\centering
\includegraphics[width=\textwidth]{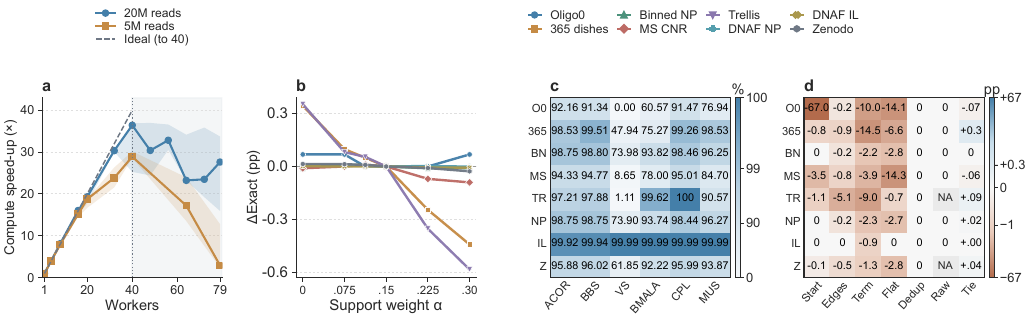}

\caption{Parallel performance, reconstruction quality and ablation effects. \textbf{a}, Compute speed-up relative to P1 medians on controlled workloads; bands transform three-run extrema. Shading marks simultaneous multithreading (SMT) beyond 40 physical cores. \textbf{b}, Paired-seed Exact changes from $\alpha=0.15$ at P20, with three-seed ranges. \textbf{c}, A800 benchmark Exact (\%) for 48 method--dataset cells (ACOR: P20); 100 denotes perfect recovery. \textbf{d}, Median paired-seed P20 ablation effects (percentage points); grey denotes incomplete output. Colour scales are asinh (c) and signed-asinh (d). Ablation labels: Start, start support; Edges, unobserved-transition admission; Term, terminal support; Flat, family priority; Dedup, duplicate-prefix merging; Raw, raw-read alternatives; Tie, tie-breaking. Ranges are descriptive min--max summaries. Panels b,d show paired-seed ablation results; c reports one A800 observation per method--dataset cell. Rows in c,d follow the eight public benchmark rows in Table~\ref{tab:benchmark-main} (Supplementary Notes~S3--S4).}
\label{fig:main-results}
\label{fig:scaling}
\label{fig:ablation-results}
\end{figure}

Million-read inputs also completed in seconds in the A800 acceptance batch (Supplementary Note~S4; Supplementary Note~\ref{sec:implementation}). DNAFormer Illumina's 3.18 million reads took 3.25~s with 99.92\% Exact, and DNAFormer nanopore's 2.35 million reads took 4.35~s with 98.75\% Exact. These were the shortest retained engine-time observations for each dataset among the tested settings, whose predictions were identical. The original benchmark timing comparison also showed seconds-scale processing of million-read libraries (Fig.~\ref{fig:runtime-all}).

These runtime comparisons use method-specific execution configurations and loading scopes, as defined under Evaluation metrics and in Supplementary Note~S5.

\subsection{Memory efficiency and workload-dependent scalability}

Hybrid storage limited high-$k$ allocation to observed keys while preserving dense-reference predictions in all 32 validated configurations. At $k=12$, peak RSS was 41.84--658.97~MiB for hybrid storage, compared with 22.53--23.14~GiB for the historical dense implementation (Fig.~\ref{fig:k-evidence}b). This corresponded to a 36--552-fold reduction in the tested configuration comparisons, which used different worker allocations. The prediction checks confirmed unchanged reconstruction outputs despite the reduction in memory use.

On a controlled 20-million-read workload, median compute time decreased from 47.2~s to 1.30~s, giving $36.4\times$ speed-up and 91.0\% parallel efficiency (Fig.~\ref{fig:main-results}a). Efficiency was 72.2\% on five million reads, indicating stronger scaling with larger workloads. Additional hardware threads beyond the physical cores provided no further compute-time benefit. Complete engine-time speed-up reached $6.46\times$, with median input loading equal to 84.4\% of median engine time. Predictions remained identical, locating the observed scaling limits in workload size and input handling rather than changes in reconstruction quality (Supplementary Note~S6).

\subsection{Ablation, context sensitivity and extended reconstruction}

Historical ablation summaries identified boundary evidence and candidate selection as major contributors to recovery (Fig.~\ref{fig:main-results}b,d). Removing start support reduced mean Exact by 9.1 percentage points; removing terminal support or using the flat-selection variant reduced it by 5.5 points. Allowing unobserved transitions reduced Exact and increased runtime, whereas raw-read alternatives prevented missing outputs in some clusters.

\begin{figure}[t]
\centering

\includegraphics[width=\textwidth]{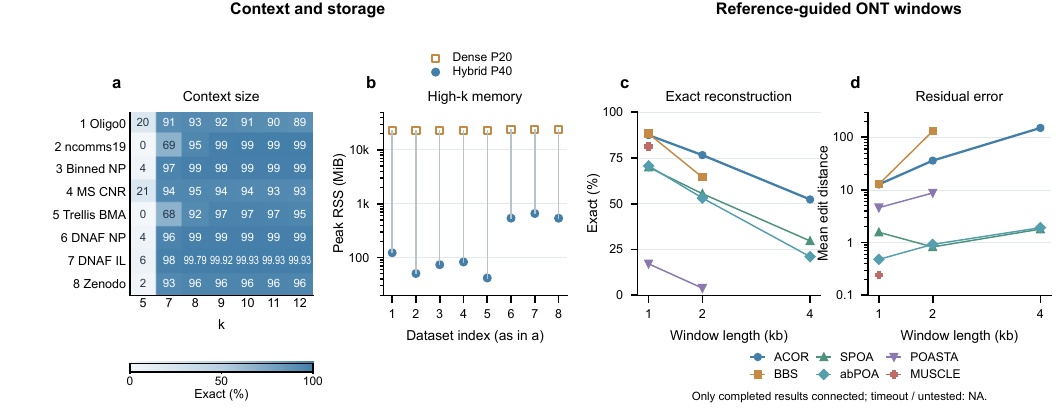}
\caption{Context length, memory and long-sequence reconstruction. \textbf{a}, Exact across the seven-setting dense $k$ grid at P20. \textbf{b}, Process peak RSS at $k=12$, comparing historical dense P20 with hybrid P40. \textbf{c,d}, Exact and Mean ED for reference-guided ONT windows at P16. The 1-, 2- and 4-kb sets contain 256, 256 and 128 regional windows, respectively. Unavailable results are labelled separately. RSS and Mean ED use logarithmic axes. Comparison protocols and stopping decisions are reported in Supplementary Notes~S6--S7.}
\label{fig:k-evidence}
\end{figure}

The seven-setting $k$ grid showed dataset-specific context preferences (Fig.~\ref{fig:k-evidence}a), and synthetic sweeps extended strand-length tests to 8~kb. A retrospective, reference-informed rule came within 0.144 percentage points of each dataset's best-tested Exact; it was not a reads-only selector (Supplementary Note~S3).

Reference-guided nanopore windows tested reconstruction over 1--4~kb (Fig.~\ref{fig:k-evidence}c,d). At 2~kb, ACOR achieved 76.56\% Exact, above the completed BBS, SPOA, abPOA and POASTA-based results. At 4~kb, Exact was 52.34\%, compared with 29.69\% for SPOA and 21.09\% for abPOA. Each window set completed in under one second in the tested configuration, with identical predictions across validated worker allocations. However, ACOR had higher Mean ED and lower Accuracy across the full 4-kb window set. Its non-exact outputs included severe reconstruction errors (Supplementary Note~S7).

\begin{table}[!tbp]
\caption{Full-library comparison on our NGS data (96,055 designs; complete outputs). ACOR used 16 workers; baselines used 16 wrapper task slots (Supplementary Note~S5).}
\label{tab:own-ngs-p16}
\centering
\fontsize{10}{12}\selectfont
\setlength{\tabcolsep}{2.5pt}
\renewcommand{\arraystretch}{1.05}
\begin{tabular}{@{}lrrrr@{}}
\hline
Method & Exact (\%) & Max ED & Time (s) & RSS (MiB) \\
\hline
ACOR ($k=9$)  & 99.9219 & 50 & 11.805     & 2775.9 \\
ACOR ($k=15$) & 99.9844 & 2  & 20.526     & 2493.0 \\
abPOA        & 99.9833 & 2  & 847.826    & NR \\
BBS          & 99.9844 & 2  & 52.786     & NR \\
BMALA        & 99.9833 & 2  & 26.407     & NR \\
CPL          & 99.9885 & 2  & 413.993    & NR \\
MUSCLE       & 99.9844 & 2  & 17,893.918 & NR \\
SPOA         & 74.1075 & 22 & 772.890    & NR \\
VS           & 99.9833 & 2  & 168.218    & NR \\
POASTA-based & 82.6526 & 19 & 830.333    & NR \\
\hline
\end{tabular}
\vspace{0.5pt}
\begin{minipage}{\linewidth}
\footnotesize
\textit{Note:}
Times are single-run observations with method-specific loading scopes. ACOR $k=9$: original comparison window; $k=15$: concurrent sensitivity tests. Max ED: maximum Levenshtein distance; RSS: whole-process peak; NR: not recorded. MUSCLE: P16 controller elapsed.
\end{minipage}
\end{table}

\section{Discussion and conclusion}
\label{sec:discussion}

ACOR provides a streaming consensus reconstruction method that combines incremental evidence accumulation with a single decoding step at cluster closure. Across the evaluated DNA-storage benchmarks, it achieved competitive reconstruction quality with seconds-scale processing of million-read inputs. Its central contribution is to incorporate evidence as reads arrive without repeatedly reconstructing the consensus sequence. This separation provides a computational basis for integrating reconstruction with ongoing data acquisition.

The broader evaluation shows that reconstruction quality, computational throughput and memory efficiency can be addressed within the same framework. Historical ablation summaries suggest that evidence-guided candidate generation and selection contribute to reconstruction quality. Sparse storage avoids the exponential allocation of dense high-$k$ counters while preserving predictions in the tested configurations. Coordinated parallel execution substantially reduces reconstruction-compute time, with workload size and input loading setting different limits on overall performance. Reference-guided nanopore experiments extend the evaluation to kilobase-length sequences, but more severe errors among failed reconstructions remain a limitation.

Our short-read NGS wet-lab library complements public benchmarks and simulations: longer contexts improved Exact and reduced maximum residual errors, supporting evaluation of both recovery and error severity.

Current evaluations use assigned clusters, known design lengths and externally determined closure. Integrating online read assignment and closure would extend ACOR to a complete live readout pipeline. The retrospective $k$-selection rule uses references; reads-only selection and stronger error control for difficult long sequences are priorities for future work.

\section{Data availability}

Benchmark data are available from Chandak nanopore and UW MISL 365-dishes \citep{chandak2019nanoporedata,uwmisl2019ncomms19data}, and Microsoft CNR and Trellis BMA \citep{microsoft2021cnrdataset,microsoft2024trellisbma}. DNAFormer and end-to-end storage data are in Zenodo records 13896773 and 10943282, respectively \citep{sabary2024dnaformerdatasets,sokolovskii2024endtoenddata}. ONT analyses used \href{https://www.ebi.ac.uk/ena/browser/view/SRR21864871}{SRR21864871} reads and chromosome \href{https://www.ncbi.nlm.nih.gov/nuccore/NZ_CP024549.1}{NZ\_CP024549.1} from assembly GCF\_002753405.1. Source Data comprise the full own-NGS assigned-cluster bank, design references, retained results and protocols, with a figure/table index. The archive excludes raw paired-end NGS files and third-party raw reads and reference sequences.

\section{Code availability}
 ACOR (v0.2.0) is publicly available under the MIT License at:
\noindent\url{https://github.com/Jarvis5272/ACOR-Real-Time/releases/tag/v0.2.0}. The release is permanently archived at \href{https://archive.softwareheritage.org/swh:1:rev:bcc07ca3d55151aa55fd86a9e4509dadc8240761}{Software Heritage}.
The release includes source code, documentation and tests, with defaults in \texttt{provenance/ALGORITHM\_LOCK.json}. Research-build settings, timings and experimental variants are documented in the Supplementary information and Source Data.

\section*{Funding}
This work is supported by the National Key Research and Development Program of China (2025YFC3409900), the Tianjin Municipal Science and Technology Major Program (25ZXZSSS00390), the Tianjin Future Industry
Science and Technology Major Project (25ZXWCSY00290),
the Emerging Frontiers Cultivation Program of Tianjin
University Interdisciplinary Center, and the Independent
Research Project of State Key Laboratory of Synthetic
Biology (HCZC-202605A).

\section*{Supplementary data}
Supplementary information follows the main text and references in this preprint.\\
Conflict of interest: None declared.

\FloatBarrier
\bibliographystyle{oup-abbrvnat}
\bibliography{references}
\ACORStartSupplement
\begin{center}
{\Large\bfseries Supplementary information}\par\medskip
{\large ACOR: incremental evidence accumulation towards real-time DNA storage reconstruction}\par\medskip
Linxuan Han, Huaming Wu, Shi Jia and Guanjin Qu
\end{center}

This supplement collects the completed experiments supporting ACOR: parameter and component comparisons, public and own-NGS benchmarks, memory and parallel scaling, long-sequence stress tests, complete benchmark timing records and release compatibility. Experimental settings, measurement scopes and Source Data are provided with the results.

\section{Experimental settings and measurement}
\label{sec:implementation}\label{sec:original-settings}\label{sec:ledger-semantics}

\subsection{Datasets, defaults and execution environments}
Experiments covered eight public DNA-storage benchmarks, our assigned-cluster NGS library, synthetic workloads and reference-guided Oxford Nanopore Technologies (ONT) windows. Binned nanopore overlaps DNAFormer nanopore, and Trellis BMA is simulated. Dataset and assignment protocols accompany the corresponding results.

Unless a changed factor is stated, the original benchmark used $k=9$, beam width $B=16$, support weight $\alpha=0.15$ and design-length tolerance $\delta=5$. The log-length standard-deviation floor was $\sigma_0=0.08$, spread multiplier $\tau=2$, and median-relative length multipliers $\rho_-=0.6$ and $\rho_+=1.5$. Length and terminal score weights were $\lambda_{\rm len}=0.375$ and $\lambda_{\rm end}=0.05$.

The original Linux server, LEGACY-116, had two Xeon E5-2660 v3 processors, 20 physical cores, 40 logical CPUs and approximately 125.77~GiB of OS-visible memory. A800-173 had two Xeon Silver 4316 processors, 40 physical cores, 80 logical CPUs and approximately 503~GiB of OS-visible memory. Windows was used for local preparation. The original worker scan tested $P=1,2,4,8,12,16,20,24,28,32$, with three recorded seeds per setting.

\subsection{Reconstruction settings used in the experiments}
Each run processed assigned reads and decoded once at externally signalled cluster closure, using a fixed target length per cluster. Exact length medians used retained samples or two heaps; even-sample medians averaged the two middle lengths. Only canonical ACGT runs contributed contexts; noncanonical symbols interrupted the scan.

Root contexts were ranked by decreasing start support, with occurrence count as the fallback and increasing encoded context as the tie-breaker. Beam and candidate scores accumulated at full precision. Ranking used \texttt{llround} at scale $10^{12}$, with halfway cases rounded away from zero, followed by increasing lexical sequence order. Neighbouring-length candidates at equal offsets below and above the design length were ranked together. Raw-read alternatives used decreasing normalized-string frequency and then lexical order.

\subsection{Quality, timing and memory summaries}
Quality was evaluated per assigned cluster or genomic window, using the main-text definitions of Exact, Accuracy and Mean ED. Read-order seeds and repeated timings represent technical variation. Repeated observations use the stated median and observed minimum--maximum range; single records are reported as observed values. Missing or stopped configurations are marked unavailable; measured zero values are retained.

Runner wall time includes ACOR input/order loading, reconstruction and prediction writing, excluding external preparation and evaluation. Reconstruction-compute time excludes input/order loading and prediction writing but retains initialization, scheduling and reconstruction. Historical files sometimes call the broader runner interval ``engine time''. Baseline native durations, wrapper elapsed times and sums of per-cluster times retain their recorded scopes and worker allocations. Peak RSS denotes process resident-set high-water memory; sampled parent-plus-child baseline RSS is identified where used. Linux record suffix \src{_kb} denotes KiB.

Main Fig.~2 uses ACOR three-seed runner-time medians and minimum--maximum ranges. The original baselines used one native BBS or BMALA process, serial VS, eight CPL shards or sixteen MUSCLE shards. Original CPL quality uses saved first-run predictions; MUSCLE 5.3 has one completed diagnostic measurement per dataset. The separate A800 quality batch and own-NGS comparison are reported in Notes~\ref{sec:public-quality} and~\ref{sec:own-ngs}.

\subsection{Measured builds and baseline adapters}
The original benchmark, controlled-scaling, high-context and release comparisons retain their measured build identities. The hybrid compatibility guard passed the C++ ledger-reduction test and all 12 Python tests; an oligo0/$k=12$/P1 check retained the historical prediction. The complete matrices predate that guard. Source, binary, protocol and input identifiers are in \src{SourceData/provenance/experiment_build_identities.tsv}.

Baselines were BBS \citep{gu2025bbs}, VS \citep{viswanathan2008improved,sabary2024reconstruction}, BMALA \citep{gopalan2018trace,sabary2024reconstruction}, the native Deep-DNA consensus component CPL \citep{barlev2025scalable}, and MUSCLE consensus \citep{edgar2004muscle,edgar2022muscle5}. Long-window adapters used SPOA 4.1.5, abPOA 1.5.7, POASTA 0.1.0, MUSCLE 5.3 and minimap2 v2.30.

\begingroup\raggedright
SPOA used \texttt{\detokenize{spoa -l 1 -r 0 reads.fasta}} and abPOA used \texttt{\detokenize{abpoa -r 0 reads.fasta}}; adapters parsed their FASTA consensus output. POASTA used \texttt{\detokenize{poasta align -O fasta -o aligned.fa reads.fa}}, and MUSCLE used \texttt{\detokenize{muscle -align reads.fa -output aligned.fa -threads 1}}. Their alignments were converted to consensus by column voting \citep{antkowiak2020lowcost}: nucleotides had unit weight and gaps had weight 0.4; ties followed A, C, G, T, N, gap. Winning gaps were omitted, and other symbols were counted as N. POASTA results therefore describe this alignment-plus-consensus pipeline.\par
\endgroup

BBS used \texttt{\detokenize{bbs-local -i clusters.txt -l L -s ===}}. Shards began with the delimiter \src{===}; ordered prediction lines were mapped to the corresponding ordered clusters. SPOA/abPOA window tasks used a pool of size $P$ with CPU affinity restricted to the first $P$ logical CPUs. MUSCLE used one thread per task; BBS used $P$ independent shards. Long-window construction and method-specific timing scopes are given in Note~\ref{sec:ont-boundaries}.

\section{Validation of reconstruction decisions and parameter sensitivity}
\label{sec:design-validation}

These experiments tested component ablations, beam width, context length and synthetic read conditions. The retrospective context-selection check is reported separately from the measured reconstruction grids.

\subsection{Structural comparisons and output completeness}

Main Fig.~3b,d report support-weight and structural-ablation comparisons of reconstruction quality and output completeness. Their source records are described in Note~\ref{sec:historical-records}.

Structural ablations changed start evidence, unobserved-transition admission, terminal evidence, candidate-family priority, duplicate-prefix merging, raw-read alternatives or deterministic tie-breaking. Main Fig.~3d shows per-dataset medians of within-seed Exact differences at P20, in percentage points. Mean effects weight datasets equally; runtime ratios use within-dataset summaries followed by the cross-dataset median. The terminal-support variant and remaining variants used successive batches, so cross-batch runtime differences are descriptive.

Removing start support reduced mean Exact by 9.1 percentage points. Removing terminal support or using the flat-selection variant reduced it by approximately 5.5 points. The flat variant jointly replaced family-priority selection and its candidate handling with an empirical-length cache and global raw-read mode. Permitting unobserved transitions reduced mean Exact by approximately one point and gave a median runtime ratio of 2.34.

Removing raw-read alternatives left four Trellis BMA clusters and nine Zenodo clusters without predictions in every recorded seed. These incomplete cells are marked unavailable. Neighbouring length settings sometimes changed individual predictions while retaining aggregate Exact, so prediction identity and quality were checked separately.

\subsection{Beam-width and support-weight diagnostics}
\label{sec:beam-diagnostics}

Beam width was varied at fixed $k=9$, log-length penalty configuration and 20 workers. Widths were 2, 4, 8, 16, 24, 32, 48 and 64 on each of the eight datasets, giving 64 single-run observations. Quality differences and runtime ratios use each dataset's $B=16$ observation as the reference. Wider beams do not guarantee higher quality under the selection hierarchy.

\begin{figure}[ht!]\centering
\includegraphics[width=\textwidth]{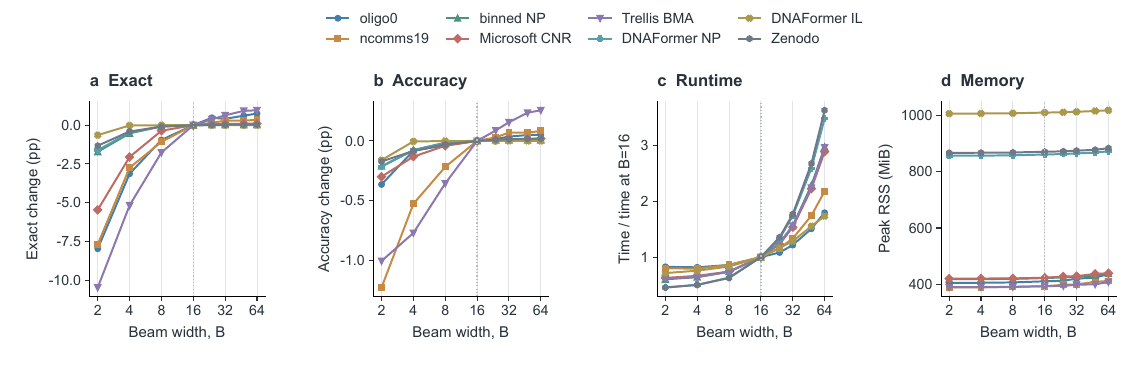}
\caption{Beam-width trade-offs at fixed worker count. \textbf{a,b}, Exact and Accuracy differences in percentage points relative to $B=16$. \textbf{c}, Wall-time ratio relative to $B=16$. \textbf{d}, Peak RSS. All measured widths, including 24 and 48, are plotted on a base-2 logarithmic horizontal axis. The dotted reference marks $B=16$. Points are single-run observations. Source Data: \texttt{beam\_P20.tsv}.}
\label{fig:beam}
\end{figure}

The $B=4$ diagnostic reduced mean Exact by 1.8 points and had a median runtime ratio of 0.74 relative to $B=16$. Note~\ref{sec:historical-records} reports three additional scoring variants measured without a matched default arm.

Main Fig.~3b reports median within-seed Exact differences from $\alpha=0.15$ at P20, with the observed three-seed range. All curves are zero at the reference weight, $\alpha=0.15$.

\subsection{Benchmark sensitivity to \texorpdfstring{$k$}{k} and measured memory}
\label{sec:benchmark-k-diagnostics}

Across eight benchmarks, the context-size grid used $k\in\{5,7,8,9,10,11,12\}$. It measured quality, runtime and memory at fixed $B=16$, log-length penalty configuration and 20 workers. All 56 configurations completed with matching reference, predicted and evaluated cluster counts. Edit-distance evaluation joined predictions and references by canonical cluster identifiers.

The grid combines 40 saved configurations at $k=5,7,9,11,12$ with 16 subsequent runs at $k=8,10$. Saved predictions were rescored after correcting the identifier join; reconstruction, timing and RSS records retain their original batches. Each configuration has one timing and RSS observation. Lines connect settings within a dataset, and runtime comparisons describe the recorded batches.

\begin{figure}[ht!]\centering
\includegraphics[width=\textwidth]{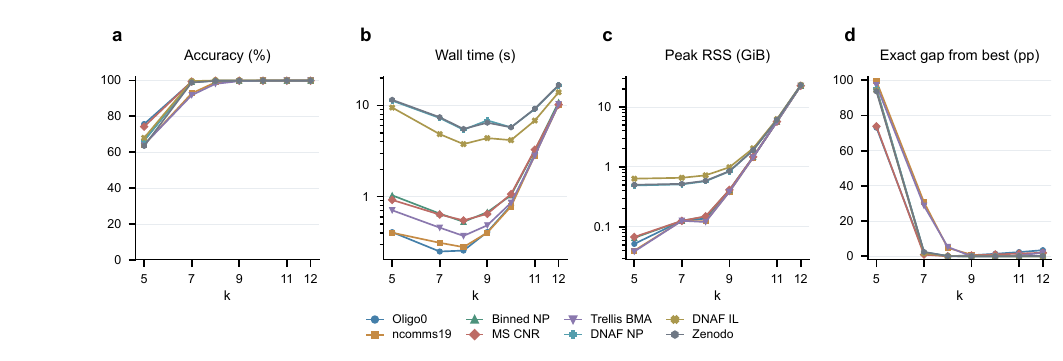}
\caption{Complete benchmark context-size grid. \textbf{a}, Accuracy. \textbf{b}, Recorded wall time. \textbf{c}, Process peak RSS. \textbf{d}, Exact shortfall from the highest tested Exact within each dataset, in percentage points. All 56 single-configuration records are shown. Colours and markers identify datasets; NP and IL denote nanopore and Illumina. Panels b and c use logarithmic vertical axes. Panel d uses each dataset's retrospective best-tested setting. Source Data: \texttt{real\_k\_grid\_full\_56.tsv}.}
\label{fig:real}
\end{figure}

Oligo0 and Microsoft CNR reached their highest tested Exact at $k=8$, exceeding $k=9$ by 0.614 and 0.451 percentage points, respectively. For ncomms19, Exact rose from 98.528\% at $k=9$ to 99.460\% at $k=12$. Trellis BMA and Zenodo reached their highest tested Exact at $k=9$.

Across the eight datasets, recorded RSS ranged from 0.384--0.987~GiB at $k=9$ to 22.532--23.135~GiB at $k=12$. Process RSS includes input storage and concurrent worker states. This benchmark grid used P20; the separate ONT-window grid used P1 on different inputs.

\subsection{Synthetic length, read-count and error-budget sweep}
\label{sec:synthetic-diagnostics}

The synthetic grid varied strand length, read count and generator error budget. Best-tested-$k$ summaries describe the highest Exact within the evaluated parameter range.

The completed grid comprises five design lengths (500, 1,000, 2,000, 4,000 and 8,000~nt), three read counts per cluster ($N=20,80,320$), three generator error budgets ($\eta=0.025,0.05,0.10$) and two seeds (3401 and 3402). Each of the 90 conditions contains eight reference strands and is run at $k\in\{5,7,9,11,12\}$, yielding 450 records. Runs use one worker, $B=16$ and the log-length penalty configuration.

Reference bases were drawn independently and uniformly from ACGT, keyed by design length and seed. Read generation additionally used read count, error budget, cluster and read identifier. Each reference base underwent deletion with probability $0.12\eta$; deletion advanced to the next position. Otherwise, replacement drew uniformly from ACGT with probability $0.50\eta$, allowing the original base to be redrawn. Insertion then occurred with probability $0.15\eta$. Here, $\eta$ specifies the injection budget rather than the measured Levenshtein error rate. References were used only for external evaluation.

All 450 runs completed and returned predictions for their eight clusters. Conditions shared reference sequences, and each configuration had one timing observation. Peak RSS is the process high-water mark, converted from recorded kilobytes to MiB by division by 1,024.

\begin{figure}[H]\centering
\includegraphics[width=\textwidth]{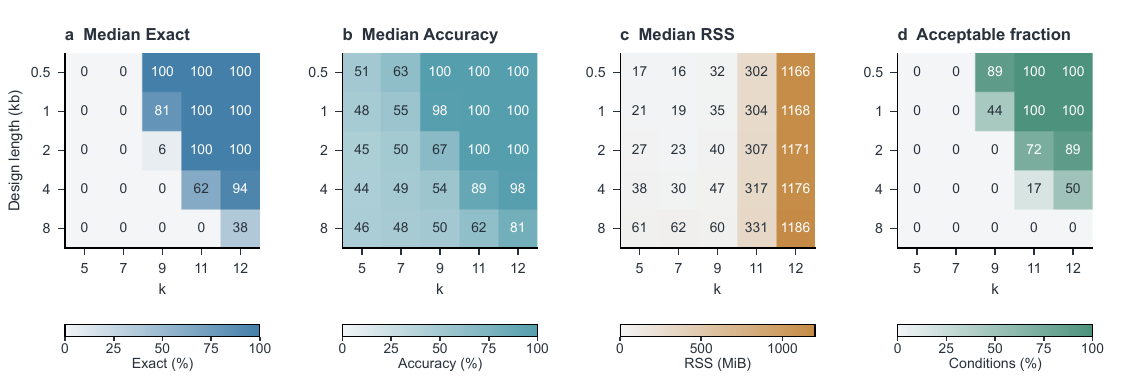}
\caption{Complete synthetic $k$ grid. \textbf{a,b}, Median Exact and Accuracy. \textbf{c}, Median peak RSS. \textbf{d}, Fraction of conditions with Exact at least 0.90. Each cell summarizes the 18 completed read-count--error-budget--seed combinations at one length and $k$. Exact and Accuracy are percentages. Source Data: \texttt{synthetic\_k\_runs.tsv} and \texttt{synthetic\_cells.tsv}.}
\label{fig:synthetic}
\end{figure}

For each condition, the best-tested-$k$ envelope takes the maximum Exact over the five tested values. At Exact $\geq0.90$, acceptable-condition counts were 18/18 at 500~nt, 18/18 at 1,000~nt, 16/18 at 2,000~nt, 9/18 at 4,000~nt and 0/18 at 8,000~nt. With eight clusters per condition, this threshold requires all eight to be exact. At 8,000~nt, $N=320$, $\eta=0.025$ and $k=12$, the two seeds recovered 0.750 and 0.875 of the strands exactly. These simulated-strand tests complement the biological long-sequence experiments in Note~\ref{sec:ont-boundaries}.

\subsection{Retrospective context-selection experiment}
\label{sec:k-rule}

This experiment compared reference-informed context choices with the highest Exact in the seven-setting benchmark grid. Selection used reference-derived within-strand collision counts, median cluster read count, reference-normalized read errors and measured process peak RSS. Profiling ordered clusters by the SHA-256 hash of their identifiers and retained at most 5,000 clusters per dataset. The first two encountered reads per retained cluster were compared with the supplied reference.

The rule selected the smallest tested context meeting a mean collision count at most 0.30, estimated support at least 6.5 and peak RSS of at most 2.5~GiB. Settings were $P=20$ and $B=16$. Support used a heuristic intact-context estimate from sampled read errors; RSS came from completed dense-grid runs. Thresholds were set and evaluated on the same eight datasets using reference-informed inputs.

The selected values were $k=8,10,8,8,9,9,9,8$ for Oligo0, ncomms19, binned nanopore, Microsoft CNR, Trellis BMA, DNAFormer nanopore, DNAFormer Illumina and Zenodo, respectively. All choices were within 0.2 percentage points of the highest tested Exact, with a maximum shortfall of 0.1438 points on Zenodo.

\begin{figure}[H]\centering
\includegraphics[width=\textwidth]{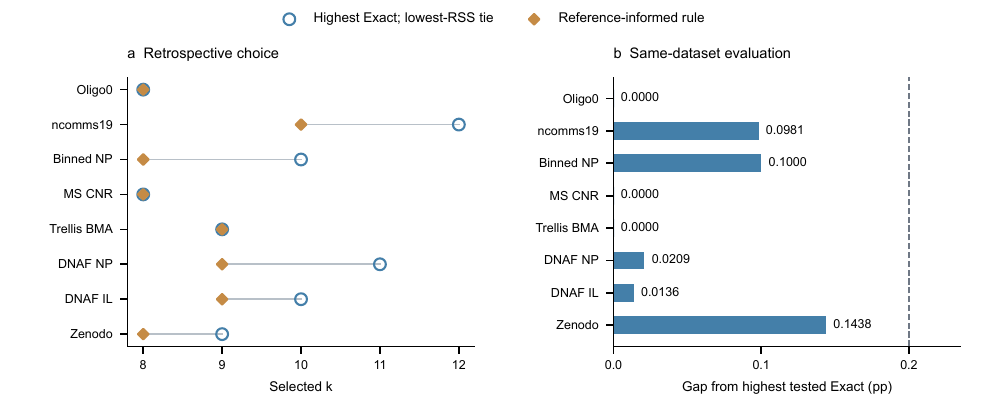}
\caption{Retrospective context-selection experiment on the complete benchmark grid. \textbf{a}, Reference-informed context choices and settings attaining the highest measured Exact, using the lowest recorded RSS to break equal-Exact ties. \textbf{b}, Exact shortfall from each dataset's highest measured Exact. The dashed line marks the descriptive 0.2-percentage-point tolerance. All eight datasets are shown. Selection used supplied references, reference-based error profiles and measured grid RSS. Source Data: \src{direct_k_selector_validation.tsv}, \src{real_dataset_profiles.tsv} and \src{real_truth_kmer_structure.tsv}.}
\label{fig:selector}
\end{figure}

The ONT experiments in Note~\ref{sec:ont-boundaries} used a separate preprocessing protocol and context settings.

\section{Detailed quality evaluation on public DNA-storage benchmarks}
\label{sec:public-quality}

The experimental inputs derive from the published nanopore-storage, CNR, Trellis BMA, DNAFormer and end-to-end-storage sources \citep{lopez2019dnaassembly,chandak2020overcoming,microsoft2021cnrdataset,srinivasavaradhan2021trellisbma,barlev2025scalable,welter2026endtoend}. The corresponding public data records are \citep{chandak2019nanoporedata,uwmisl2019ncomms19data,microsoft2024trellisbma,sabary2024dnaformerdatasets,sokolovskii2024endtoenddata}. Dataset dimensions and overlap are listed in main Table~1.

Table~\ref{tab:public-quality} contains 48 A800-173 method--dataset cells for eight datasets, ACOR and five baselines (BBS, VS, BMALA, CPL and MUSCLE). These records also supply main Fig.~3c. ACOR default prediction hashes and all three quality metrics agreed across the five recorded A800 worker settings. Worker timings and the separate historical batch are in Note~\ref{sec:historical-records}.

Each cell represents one completed observation. CPL used Deep-DNA's native classical consensus component, limited to 32 reads per cluster. Its DNAFormer nanopore result came from a full-input run with isolated invocation directories and complete native-label checks. Other cells use the saved benchmark predictions. References were used only for external quality evaluation.

Per-cell identities and quality checks are in \src{SourceData/data/final_presentation/Fig3c_record_checks.tsv}; full-input CPL predictions and invocation records are in \src{SourceData/cpl_verified_20261013/}.

Exact and Accuracy are fractions; Mean ED is the average edit count.

ACOR had the highest Exact on Oligo0, while CPL had higher Accuracy and lower Mean ED. Several baselines had higher Exact than ACOR on simulated Trellis BMA.

\begingroup
\fontsize{10}{12}\selectfont
\setlength{\tabcolsep}{4pt}
\renewcommand{\arraystretch}{1.02}
\setlength{\LTcapwidth}{\linewidth}
\begin{longtable}{@{}p{43mm}lrrr@{}}
\caption{Complete A800-173 quality matrix on the eight public DNA-storage benchmarks. Each method--dataset cell is one retained observation. CPL denotes Deep-DNA's native classical consensus component.}\label{tab:public-quality}\\
\toprule
Dataset & Method & Exact & Accuracy & Mean ED\\
\midrule
\endfirsthead
\multicolumn{5}{l}{\small\textbf{Table~\thetable\ (continued)}}\\
\toprule
Dataset & Method & Exact & Accuracy & Mean ED\\
\midrule
\endhead
\midrule
\multicolumn{5}{r}{\footnotesize Continued on next page}\\
\endfoot
\bottomrule
\endlastfoot
Oligo0 & ACOR & 0.921555 & 0.997373 & 0.283765\\
Oligo0 & BBS & 0.913370 & 0.997379 & 0.283083\\
Oligo0 & VS & 0.000000 & 0.784479 & 23.276262\\
Oligo0 & BMALA & 0.605730 & 0.985404 & 1.576398\\
Oligo0 & CPL & 0.914734 & 0.998187 & 0.195771\\
Oligo0 & MUSCLE & 0.769441 & 0.996608 & 0.366303\\
\addlinespace[4pt]
ncomms19 365 dishes & ACOR & 0.985280 & 0.997622 & 0.356722\\
ncomms19 365 dishes & BBS & 0.995093 & 0.999748 & 0.037782\\
ncomms19 365 dishes & VS & 0.479392 & 0.963376 & 5.493621\\
ncomms19 365 dishes & BMALA & 0.752699 & 0.992692 & 1.096173\\
ncomms19 365 dishes & CPL & 0.992640 & 0.999630 & 0.055447\\
ncomms19 365 dishes & MUSCLE & 0.985280 & 0.999558 & 0.066241\\
\addlinespace[4pt]
Binned nanopore & ACOR & 0.987500 & 0.998264 & 0.243100\\
Binned nanopore & BBS & 0.988000 & 0.998529 & 0.205900\\
Binned nanopore & VS & 0.739800 & 0.987242 & 1.786100\\
Binned nanopore & BMALA & 0.938200 & 0.997504 & 0.349400\\
Binned nanopore & CPL & 0.984600 & 0.998741 & 0.176300\\
Binned nanopore & MUSCLE & 0.962500 & 0.998160 & 0.257600\\
\addlinespace[4pt]
Microsoft CNR & ACOR & 0.943309 & 0.998303 & 0.186699\\
Microsoft CNR & BBS & 0.947716 & 0.998469 & 0.168369\\
Microsoft CNR & VS & 0.086538 & 0.928032 & 7.916466\\
Microsoft CNR & BMALA & 0.780048 & 0.993643 & 0.699219\\
Microsoft CNR & CPL & 0.950120 & 0.998617 & 0.152143\\
Microsoft CNR & MUSCLE & 0.846955 & 0.997642 & 0.259415\\
\addlinespace[4pt]
Trellis BMA & ACOR & 0.972060 & 0.995559 & 0.488539\\
Trellis BMA & BBS & 0.978842 & 0.999529 & 0.051811\\
Trellis BMA & VS & 0.011122 & 0.924194 & 8.338668\\
Trellis BMA & BMALA & 0.996202 & 0.999921 & 0.008680\\
Trellis BMA & CPL & 1.000000 & 1.000000 & 0.000000\\
Trellis BMA & MUSCLE & 0.905737 & 0.999090 & 0.100095\\
\addlinespace[4pt]
DNAFormer nanopore & ACOR & 0.987525 & 0.998602 & 0.195706\\
DNAFormer nanopore & BBS & 0.987543 & 0.998744 & 0.175902\\
DNAFormer nanopore & VS & 0.738979 & 0.987160 & 1.797601\\
DNAFormer nanopore & BMALA & 0.937359 & 0.997736 & 0.317024\\
DNAFormer nanopore & CPL & 0.984388 & 0.998725 & 0.178566\\
DNAFormer nanopore & MUSCLE & 0.962719 & 0.998383 & 0.226440\\
\addlinespace[4pt]
DNAFormer Illumina & ACOR & 0.999154 & 0.999902 & 0.013734\\
DNAFormer Illumina & BBS & 0.999354 & 0.999924 & 0.010678\\
DNAFormer Illumina & VS & 0.999909 & 0.999998 & 0.000309\\
DNAFormer Illumina & BMALA & 0.999927 & 0.999998 & 0.000327\\
DNAFormer Illumina & CPL & 0.999891 & 0.999998 & 0.000337\\
DNAFormer Illumina & MUSCLE & 0.999927 & 0.999998 & 0.000300\\
\addlinespace[4pt]
Zenodo 10943282 & ACOR & 0.958830 & 0.998542 & 0.218738\\
Zenodo 10943282 & BBS & 0.960202 & 0.998751 & 0.187394\\
Zenodo 10943282 & VS & 0.618525 & 0.984178 & 2.373285\\
Zenodo 10943282 & BMALA & 0.922183 & 0.997243 & 0.413543\\
Zenodo 10943282 & CPL & 0.959895 & 0.998613 & 0.208001\\
Zenodo 10943282 & MUSCLE & 0.938717 & 0.998604 & 0.209428\\
\end{longtable}
\endgroup

\section{Full-library validation on our short-read NGS data}
\label{sec:own-ngs}

The own-NGS comparison evaluated the complete pool of 96,055 designed strands.

\noindent\textbf{Own short-read next-generation sequencing (NGS) library and assignment protocol.}
Sequencing metadata reported 66,127,110 paired-end 150-nt read pairs, 65,893,682 merged inserts and 18,191,190 trimmed observations. Processing metadata specified fastp 0.24.0 for merging and Cutadapt 5.2 for trimming. Raw-pair and merge counts follow those metadata; original invocation logs and QC reports were unavailable. Assignment retained 17,701,449 observations across 96,055 distinct 100-nt designs. Read depth ranged from 3--567 per design (median, 179; 5th--95th percentiles, 86--302).

Assignment used the recorded minimap2 2.30-r1287 protocol. Filters required mapping quality at least 20, reference span at least 80~nt, aligned-query fraction at least 0.8 and full-sequence Levenshtein distance at most 20. Reads retained their reference orientation without correction or padding. References guided assignment, filtering and external evaluation; reconstruction kernels received assigned reads, design identifiers and the known 100-nt length, with reference sequences withheld. All qualifying observations were retained without read subsampling or unique molecular identifier (UMI) deduplication. Coverage therefore counts read observations, with unique molecule counts unmeasured.

Main Table~3 reports the saved P16 configurations for this library. ACOR used 16 reconstruction workers; baselines used 16 wrapper-defined task slots with method-specific native threading. Implementations included abPOA \citep{gao2021abpoa}, SPOA \citep{vaser2017racon} and a POASTA-based alignment-plus-consensus pipeline \citep{vandijk2025poasta}. All methods returned outputs for 96,055 designs; saved predictions were rescored once per configuration while retaining the original reconstruction runs.

Own-NGS CPL sampled at most 32 reads per cluster. Its adapter exposed no seed control, and quality was evaluated on one saved realization.

Each elapsed time is one observation. ACOR runner wall time includes loading and output; baseline timing starts after shared loading, and both exclude external evaluation. Duration comparisons retain these loading boundaries and recorded execution windows. ACOR RSS is whole-process peak; NR marks unrecorded baseline whole-process RSS. Maximum ED is the largest per-design Levenshtein distance. MUSCLE reports P16 controller time; a full-library P1 measurement was unavailable. P1 records used one reconstruction worker plus a separate control CPU.

Protocol and saved-run records are in \src{data/ngs_records/ngs_metadata_protocol.tsv}, \src{data/ngs_records/ngs_processing_closure.tsv} and \src{data/ngs_records/ngs_completed_quality_and_timing.tsv}.

ACOR at $k=9$/P40 recorded 9.013~s runner wall time across the complete library. Seven native single-worker configurations covered abPOA, POASTA-based, BBS, BMALA, CPL, VS and SPOA. From unrounded records, configuration-duration ratios ranged from $27.73\times$ for BMALA (249.910~s) to $1,332.73\times$ for POASTA-based reconstruction (12,011.994~s). These compare the measured P40 and single-worker configurations under their respective loading boundaries; MUSCLE P1 is excluded.

\subsection{Context-length comparison on the complete design pool}
Both $k=9$ and $k=15$ produced outputs for every design. The higher-context configuration reduced non-exact designs and the largest residual error, and its predictions matched BBS across the full pool. The public-benchmark default remains $k=9$.

Table~\ref{tab:own-ngs-context-summary} summarizes the context comparison; main Table~3 gives the full method comparison, elapsed times and peak RSS. The $k=9$/P16 timing comes from the original comparison window; $k=15$/P16 was timed during concurrent sensitivity tests.

\begin{table}[htbp]\centering\fontsize{10}{12}\selectfont
\caption{Context-length comparison on the complete own-NGS pool. Both configurations produced outputs for all 96,055 designs.}
\label{tab:own-ngs-context-summary}
\begin{tabular}{rrrr}\toprule
$k$ & Exact (\%) & Non-exact designs & Maximum ED\\\midrule
9 & 99.9219 & 75 & 50\\
15 & 99.9844 & 15 & 2\\
\bottomrule
\end{tabular}
\end{table}

\section{Prediction preservation, storage and computational scalability}
\label{sec:scaling}

\begingroup\raggedright
We tested prediction preservation, memory use and CPU scaling. Peak RSS, reconstruction-compute time and runner wall time were measured separately.\par
\endgroup

\subsection{Prediction-preserving high-context storage}
The research implementation combines a persistent worker pool, reusable worker contexts and CPU affinity with dense counters for $k\leq11$ and observed-key sparse storage for $k\geq12$. For split large clusters, high-$k$ workers accumulate private ledgers and use deterministic tree merging. The supported sparse path rejects incompatible dense shared-write routes and preserves search scores, candidate priority and tie-breaking.

The matrix covered eight benchmarks, two context settings per benchmark (the retrospective rule setting and $k=12$), and P1/P40, giving 32 configurations. Every run consumed all reads and matched its historical dense prediction SHA-256, including each worker pair. The default-quality comparison retained its original $k$ settings.

At $k=12$, summed reconstruction-compute time was 293.743603936~s at P1 and 8.567062081~s at P40, giving $34.29\times$ speed-up. Each configuration was timed once; only the P1 and P40 endpoints were measured.

\subsection{Memory comparison}
The historical dense reference used P20; the hybrid measurements used P1 and P40. All entries are whole-process peak RSS, including input/task memory, converted from Linux KiB to MiB by division by 1024. Dense P20 values spanned 22.53--23.14~GiB; hybrid P40 values spanned 41.84--658.97~MiB. These comparisons include both representation and worker-allocation differences.

\begin{table}[H]\centering\fontsize{10}{12}\selectfont
\caption{High-$k$ full-benchmark comparison at $k=12$. All predictions matched the dense reference. Reconstruction-compute times are single observations; dense memory uses the historical P20 configuration.}
\label{tab:hybrid-gate}
\begin{tabular}{lrrrrr}\toprule
Dataset & Dense P20 & Hybrid P1 & Hybrid P40 & P1 compute & P40 compute\\
& RSS (GiB) & RSS (MiB) & RSS (MiB) & (s) & (s)\\\midrule
Oligo0 & 22.55 & 41.41 & 123.08 & 4.106 & 0.161\\
ncomms19 & 22.53 & 26.45 & 50.76 & 2.626 & 0.091\\
Binned nanopore & 22.56 & 56.73 & 73.87 & 9.093 & 0.253\\
Microsoft CNR & 22.56 & 58.50 & 83.04 & 8.668 & 0.250\\
Trellis BMA & 22.53 & 28.45 & 41.84 & 5.266 & 0.149\\
DNAFormer NP & 22.99 & 500.70 & 540.50 & 99.116 & 2.636\\
DNAFormer IL & 23.14 & 649.21 & 658.97 & 59.509 & 1.531\\
Zenodo & 23.00 & 508.47 & 538.29 & 105.360 & 3.497\\
\bottomrule\end{tabular}
\end{table}

Note~\ref{sec:implementation} identifies the archived validation build and subsequent oligo0/$k=12$/P1 compatibility checks.

\begin{figure}[htbp]\centering
\includegraphics[width=\textwidth]{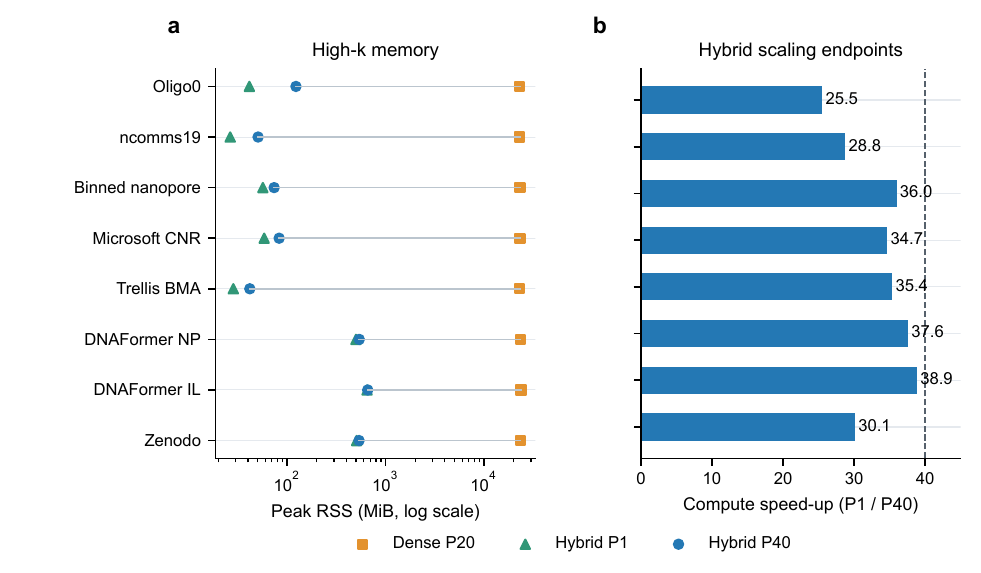}
\caption{High-$k$ storage and compute endpoints on eight benchmarks. \textbf{a}, Whole-process peak RSS for historical dense P20 and hybrid P1/P40 at $k=12$, on a logarithmic scale. The configurations differ in representation and worker count. \textbf{b}, Hybrid P1-to-P40 reconstruction-compute speed-up, with 40-fold ideal scaling. Each value is one observation. All 16 high-$k$ predictions matched the dense reference. Source Data: \src{real_k_grid_full_56.tsv} and \src{HYBRID_MATRIX.tsv}.}
\label{fig:hybrid-supp}
\end{figure}

\subsection{Execution route for the controlled scaling workloads}
The scaling workloads comprised many independent clusters and tested inter-cluster scheduling.
The optimized CPU implementation was tested on A800-173, with two Intel Xeon Silver 4316 processors and 40 physical cores in total. The 80 logical CPUs provide two hardware threads per physical core. The implementation retains one work context per worker and reuses it across assigned clusters. For inter-cluster pooling, workers are pinned before allocating their local evidence arrays, providing NUMA first-touch placement. Small and medium clusters are dynamically claimed from a shared task index. In the $k=9$ adaptive mode, clusters with at least $\max(10000,512P)$ reads instead use the intra-cluster range-merge path.

For $P\leq40$, affinity spreads workers across sockets before adding SMT siblings. The first socket receives $\lceil P/2\rceil$ workers and the second the remainder. Above 40 workers, additional workers share physical cores through SMT. The 5M and 20M workloads have 1,000 reads per cluster and test inter-cluster pooling rather than single-cluster or beam-search parallelism.

\subsection{Controlled inputs, repetitions and quality gates}
The 5M workload contains 5,000 clusters, each with 1,000 reads of length 100~nt. Reference bases are independently sampled from ACGT. Each read position is substituted with probability 0.003, choosing a different base; no insertions or deletions are introduced. The generator uses deterministic cluster/read-specific seeds. The 20M workload copies this input four times and assigns unique cluster and read identifiers, yielding 20,000 clusters. It increases computational volume without adding independent sequence diversity.

The 5M series used workers $ 1, 4, 8, 16, 20, 32, 40, 64, 79$. The 20M series added $48,56,72$, yielding 12 worker settings. Each setting was measured three times on the same fixed input. The 20M execution orders were ascending, descending, and $20,1,40,4,64,8,79,16,48,32,72,56$. Both series used $k=9$, $B=16$, and the log-length penalty configuration.

All 27 5M runs and 36 20M runs consumed every read, achieved Exact and Accuracy of 1.0, and had identical prediction-file hashes within each workload. A separate output check covered eight benchmarks at nine worker counts, producing 72 configuration records, partly reusing completed outputs. All consumed the complete input and matched the dataset-specific reference Exact, Accuracy and prediction hash.

\subsection{Timing definitions and summaries}
Reconstruction-compute time is the runner's \texttt{compute\_seconds}, covering worker start-up, state initialization, dispatch and reconstruction. It excludes initial input/order loading and prediction writing. Runner wall time includes loading and prediction writing, but excludes external order generation, quality evaluation and final metadata generation. The older benchmark's ``engine time'' uses this runner-level boundary.

Compute speed-up was the P1 median compute time divided by the median at the tested worker count. Parallel efficiency was this speed-up divided by the worker count.
In the main speed-up panel, shaded ranges divide the fixed P1 median by the maximum and minimum observed times at each $P$. Supplementary timing and RSS panels show medians with observed minima--maxima. Efficiency ranges apply the same transformation and division by $P$. The plots include every tested worker count.

\begin{figure}[H]\centering
\includegraphics[width=\textwidth]{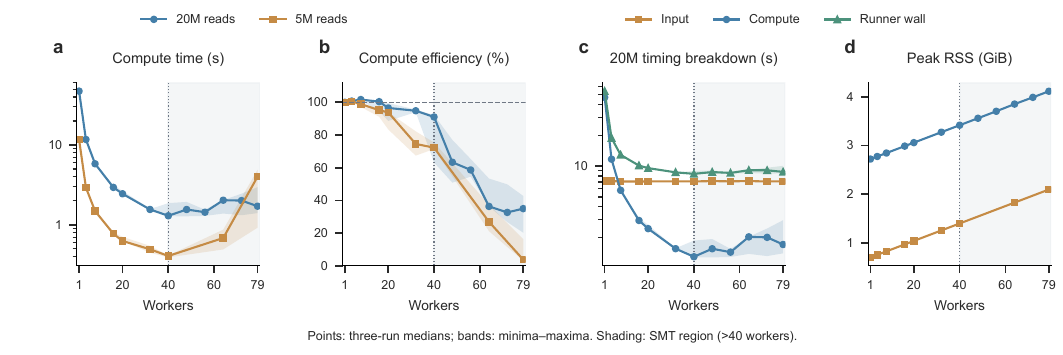}
\caption{Controlled CPU scaling and remaining input cost on A800-173. \textbf{a}, Reconstruction-compute time on 5M and 20M reads. \textbf{b}, Compute parallel efficiency. \textbf{c}, Input-loading, compute and runner wall times for 20M reads; these are separately summarized durations, not stacked components. \textbf{d}, Process peak RSS. Points are three-run medians and bands span the observed minima--maxima. The efficiency denominator uses worker count, including SMT workers above 40. Grey shading marks this SMT region; dotted lines mark 40 physical cores. Logarithmic vertical axes are used in a,c. Source Data: \texttt{parallel\_5m\_runs.tsv} and \texttt{parallel\_20m\_runs.tsv}.}
\label{fig:parallel}
\end{figure}

\subsection{Measured scaling and input cost}
On 20M reads, median reconstruction-compute time decreased from 47.203~s at P1 to 1.296~s at P40, giving $36.41\times$ speed-up and 91.0\% efficiency. It decreased at every tested step through P40. The 5M series reached $28.90\times$ and 72.2\% at P40, showing workload-dependent scaling. Some lower worker counts produced measured time ratios corresponding to efficiencies slightly above 100\%.

At P40 on 20M reads, median input time was 7.099~s and median runner wall time was 8.414~s. The ratio of these separately calculated medians was 84.4\%. Runner wall time improved $6.46\times$ from P1, with input loading accounting for most of the remaining time.

The auxiliary \texttt{parallel\_active\_seconds} subtracts measured worker start-up from reconstruction-compute time. Its P40 median was 1.227~s, giving a P1-to-P40 ratio of $38.48\times$. This elapsed interval still includes the controlling worker's post-barrier context initialization. We use the full reconstruction-compute interval, with 91.0\% efficiency, for the primary scaling result.

Above P40, this interval remained approximately 1.20--1.26~s while reconstruction-compute time increased. Median RSS rose from about 3.42~GiB at P40 to 4.11~GiB at P79, and scaling saturated in the SMT region. The recorded \texttt{cpu\_seconds\_sum} fields aggregate elapsed operation durations rather than hardware-counter CPU utilization.

Small-window worker measurements are reported in Note~\ref{sec:ont-boundaries}. Notes~\ref{sec:implementation} and~\ref{sec:historical-records} provide experimental settings and complete benchmark timing records.

\section{Long-sequence stress tests and failure boundaries}
\label{sec:ont-boundaries}

We tested longer sequences using an expanded ONT comparison and an earlier dense-storage pilot, with separate window sets.

The test used public ONT reads from accession \href{https://www.ebi.ac.uk/ena/browser/view/SRR21864871}{SRR21864871}. The reference was the \textit{Klebsiella pneumoniae} INF163 assembly GCF\_002753405.1, chromosome \href{https://www.ncbi.nlm.nih.gov/nuccore/NZ_CP024549.1}{NZ\_CP024549.1}. Reads were aligned using minimap2 v2.30 \citep{li2018minimap2}. Only primary alignments to this chromosome with mapping quality at least 30 and a CIGAR string were retained. If multiple eligible alignments remained, mapping quality and then alignment length determined the retained alignment.

\subsection{Expanded comparison: assigned inputs and window construction}
The expanded sets contained 256 windows at 1~kb, 256 at 2~kb and 128 at 4~kb, with a median of 20 retained reads per window. Reference coordinates were used to group and trim reads; all consensus callers received the same assigned reads without truth sequences. Windows were regional subsamples from one isolate and sequencing run. ACOR used $k=13,15,15$, respectively, evaluated separately from the benchmark context-selection rule.

The expanded builder examined four times as many evenly spaced candidate intervals as the requested number of windows. Candidate centres covered the chromosome between margins of $L/2+1$ bases. Candidates required at least eight spanning reads; an evenly spaced selection from the eligible candidates provided the requested windows. Within each window, eligible reads were ordered by the SHA-256 hash of the window and read identifiers. Reference-guided CIGAR projection supplied orientation and boundaries, retained insertions strictly inside the interval, and omitted deleted reference positions. Projected reads outside $[0.5L,1.5L]$ were discarded; at most 20 reads were retained, and at least eight had to survive trimming.

The final 1/2/4-kb sets contained 5,044/4,999/2,418 trimmed read records. Counts ranged from 12 to 20 per window, with a median of 20 at every length. Source reads and genomic regions could recur across windows or lengths. The saved \src{window_manifest.tsv} identifies each interval and retained read count.

Comparisons used 16 CPU workers in total: ACOR's pool or single-threaded SPOA \citep{vaser2017racon}, abPOA \citep{gao2021abpoa}, BBS \citep{gu2025bbs}, POASTA \citep{vandijk2025poasta} or MUSCLE jobs \citep{edgar2004muscle,edgar2022muscle5}. MUSCLE and POASTA alignments underwent gap-weighted column consensus extraction; POASTA results describe this alignment-plus-consensus pipeline. BBS produced exactly $N$ correctly ordered predictions; its leading delimiter gave a diagnostic count of $N+1$ boundaries.

Software versions, executable identities and adapter commands are retained in Section~\ref{sec:implementation}.

\subsection{Repetitions, timing scopes and stopping states}
Latest ACOR timings are single observations per configuration. BBS and completed POASTA settings use three timing repeats. SPOA/abPOA use three repeats at 1/2~kb and two at 4~kb; baseline times are medians of those repeats. MUSCLE has one completed 1-kb run.

SPOA/abPOA elapsed time included dataset loading, temporary input creation, consensus, prediction writing and wrapper overhead, but excluded evaluation. BBS/POASTA/MUSCLE timing included temporary inputs and consensus extraction, but excluded dataset loading, final prediction-file writing and evaluation. These wrapper-specific intervals differ from ACOR's runner wall time.

Baseline memory was the maximum summed RSS of the Python parent and descendants, sampled approximately every 10~ms and summarized by the largest value across repeats. ACOR used its runner-recorded peak RSS. The wrappers specified a 120-s window timeout and, for BBS, a 600-s shard timeout. The stopping record distinguishes the unrun SPOA 4-kb/P1 setting from observed timeouts. Runtime and memory therefore retain method-specific measurement boundaries.

MUSCLE 2~kb/P16 and POASTA 2~kb/P1 were stopped at the predeclared 600-s limit; their incomplete runs received no quality values. POASTA 2~kb/P16 completed three repetitions. Source Data identify stopping and compatibility states for omitted 4-kb settings. This protocol covered the listed adapters, excluding CPL and storage-code-dependent methods.

\subsection{Complete expanded quality and runtime comparison}
\begin{table}[htbp]\centering\fontsize{10}{12}\selectfont
\caption{Expanded ONT comparison at P16. Exact and Accuracy are fractions. Elapsed times follow the method-specific wrapper boundaries stated above; repetition counts differ. NA denotes an incomplete timeout, not zero quality.}\label{tab:ont-comparison}\begin{tabular}{rlrrrrr}
\toprule
kb & Method & Exact & Accuracy & Mean ED & Elapsed (s) & RSS (MiB)\\\midrule
1 & ACOR & 0.8750 & 0.9873 & 12.742 & 0.234 & 71.91\\
1 & SPOA & 0.6992 & 0.9984 & 1.570 & 1.772 & 927.79\\
1 & abPOA & 0.7070 & 0.9995 & 0.480 & 0.755 & 149.10\\
1 & BBS & 0.8828 & 0.9871 & 12.867 & 1.024 & 123.86\\
1 & POASTA & 0.1719 & 0.9954 & 4.621 & 25.227 & 212.92\\
1 & MUSCLE 5.3 & 0.8125 & 0.9998 & 0.246 & 427.068 & 671.17\\
2 & ACOR & 0.7656 & 0.9819 & 36.145 & 0.759 & 129.31\\
2 & SPOA & 0.5547 & 0.9996 & 0.828 & 8.209 & 5154.88\\
2 & abPOA & 0.5312 & 0.9995 & 0.926 & 1.953 & 866.86\\
2 & BBS & 0.6445 & 0.9344 & 131.184 & 3.169 & 137.20\\
2 & POASTA & 0.0391 & 0.9956 & 8.766 & 103.207 & 580.50\\
2 & MUSCLE 5.3 & NA & NA & NA & $>600$ & NA\\
4 & ACOR & 0.5234 & 0.9623 & 150.805 & 0.857 & 208.90\\
4 & SPOA & 0.2969 & 0.9995 & 1.805 & 21.703 & 19922.95\\
4 & abPOA & 0.2109 & 0.9995 & 1.922 & 2.789 & 3495.39\\
\bottomrule\end{tabular}\end{table}

At 1~kb, ACOR recovered 224/256 windows exactly, below BBS's 226/256. At 2~kb, ACOR recovered 196/256, versus BBS 165/256, SPOA 142/256, abPOA 136/256 and POASTA 10/256. At 4~kb, ACOR recovered 67/128, versus SPOA 38/128 and abPOA 27/128.

\begin{figure}[htbp]\centering
\includegraphics[width=\textwidth]{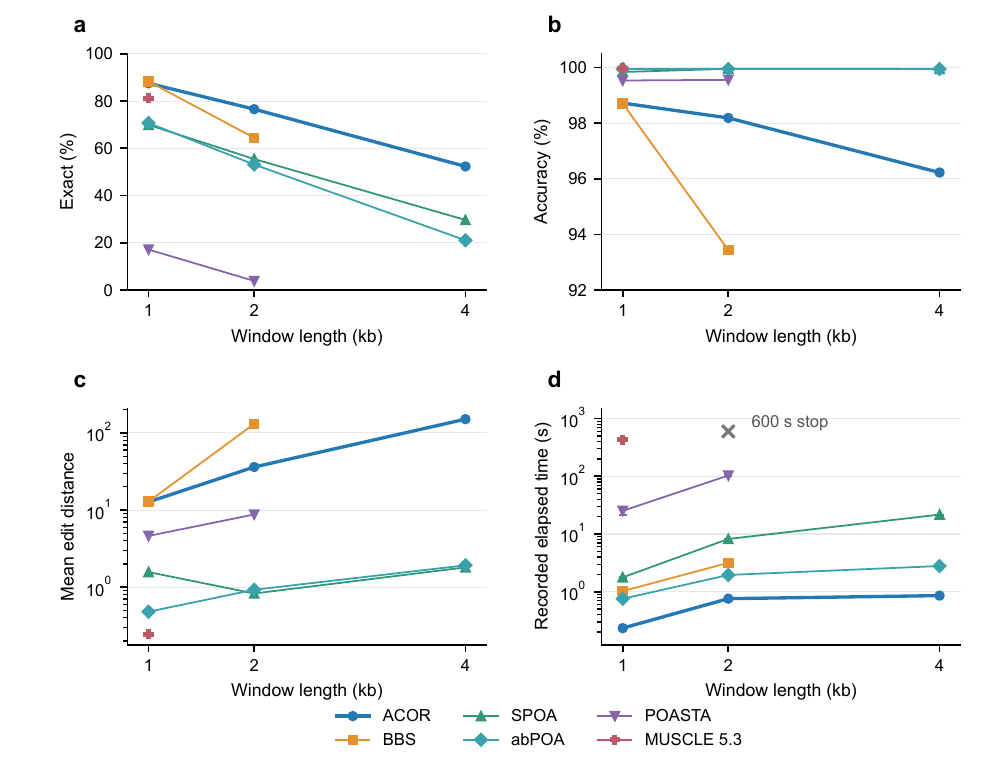}
\caption{Expanded ONT comparison. \textbf{a}, Exact. \textbf{b}, Accuracy. \textbf{c}, Mean ED. \textbf{d}, Recorded elapsed time at P16. All completed method--length cells are shown for the 256/256/128 windows. Grey crosses mark the MUSCLE 2-kb 600-s stopping threshold; other absent settings are unplotted. ACOR times are single observations; baseline times are medians with observed ranges where available. Wrapper-specific timing scopes are given in Section~\ref{sec:ont-boundaries}. Source Data: \src{REAL_ONT_LATEST_COMPARISON.tsv} and \src{ont/} replicate records.}
\label{fig:ont-expanded}
\end{figure}

\subsection{Worker dependence and prediction preservation}
\begin{table}[H]\centering\fontsize{10}{12}\selectfont
\caption{ACOR worker comparison across nine configurations. Predictions matched the earlier sparse build and were identical across worker settings at each length. Each timing is one observation.}
\label{tab:ont-workers}
\begin{tabular}{rrrrrrr}\toprule
kb & $k$ & Workers & Compute (s) & Runner wall (s) & RSS (MiB) & Exact\\\midrule
1 & 13 & 1 & 2.6266 & 2.6404 & 20.97 & 0.8750\\
1 & 13 & 16 & 0.2210 & 0.2340 & 71.91 & 0.8750\\
1 & 13 & 40 & 0.1891 & 0.2011 & 150.76 & 0.8750\\
2 & 15 & 1 & 5.9314 & 5.9487 & 29.23 & 0.7656\\
2 & 15 & 16 & 0.7412 & 0.7586 & 129.31 & 0.7656\\
2 & 15 & 40 & 1.2012 & 1.2172 & 280.33 & 0.7656\\
4 & 15 & 1 & 6.9048 & 6.9235 & 34.16 & 0.5234\\
4 & 15 & 16 & 0.8412 & 0.8566 & 208.90 & 0.5234\\
4 & 15 & 40 & 0.8837 & 0.8997 & 478.61 & 0.5234\\
\bottomrule\end{tabular}
\end{table}

All nine configurations consumed their full inputs. P40 reduced 1-kb reconstruction-compute time relative to P16, but increased it at 2/4~kb. These 128--256-cluster workloads gained less from additional workers than the controlled 20M workload.

\subsection{Error severity in failed expanded-window reconstructions}
ACOR Mean ED at 1/2/4~kb was 12.742/36.145/150.805, and maximum ED was 424/887/2926. At 4~kb, Accuracy was 0.962299, compared with 0.999549/0.999520 for SPOA/abPOA. ACOR recovered more windows exactly at 2/4~kb, but some failed outputs had large errors.

\subsection{Historical dense-storage pilot}
\label{sec:ont-pilot}
\subsubsection{Pilot window construction and sampling}
For each length $L\in\{1000,2000,4000,8000,16000,32000\}$, 12 window centres were placed at evenly spaced chromosome fractions from 0.05 to 0.95. Centres were rounded to the nearest 1,000 bases. Reads had to span the entire reference window. Their orientation and boundaries were normalized by reference-guided CIGAR projection. Inserted read bases strictly inside the window were retained; deleted reference positions contributed no read bases.

Eligible reads were ordered deterministically by a hash of window and read identifiers. At most 20 reads were selected per window. Projected reads outside $[0.5L,1.5L]$ were discarded. Windows required at least eight retained reads for lengths 1--8~kb, six for 16~kb, and three for 32~kb. References supplied mapping, grouping and trimming upstream and evaluation downstream; reconstruction kernels received only the assigned reads. The experiment measures consensus reconstruction of reference-guided windows.

The retained sets comprised 12 windows at each of 1, 2, 4 and 8~kb, ten at 16~kb and eight at 32~kb. Median retained read counts were 20, 20, 20, 15.5, 8 and 3.5, respectively. Different lengths reused centres and could contain nested regions, with coverage and sequence content also changing across lengths.

\subsubsection{Completed grid and observed limits}
\begin{figure}[H]\centering
\includegraphics[width=\textwidth]{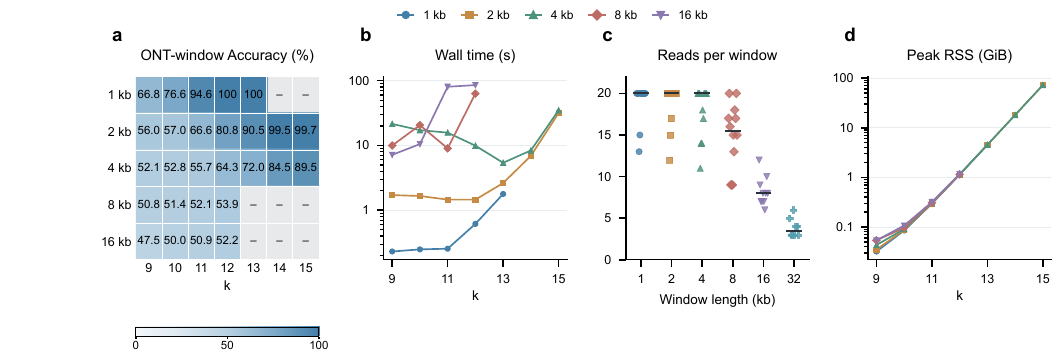}
\caption{Reference-guided ONT-window measurements. \textbf{a}, Accuracy for 27 completed length--$k$ combinations; grey cells with dashes are untested. \textbf{b,d}, Recorded wall time and peak RSS, with logarithmic vertical axes. Colours and markers identify window length. \textbf{c}, Retained read counts, including eight 32-kb windows without completed quality records. Horizontal offsets separate points; bars mark medians. Windows came from one isolate and sequencing run. Source Data: \texttt{real\_bio\_k\_grid.tsv} and \texttt{real\_bio\_window\_manifest.tsv}.}
\label{fig:bio}
\end{figure}

The 27 completed configurations used one worker, $B=16$ and the log-length penalty configuration. The tested values were $k=9$--13 at 1~kb, $k=9$--15 at 2 and 4~kb, and $k=9$--12 at 8 and 16~kb. Each configuration was recorded once. The 32-kb attempt produced no completed quality record; only its retained read counts are shown in Fig.~\ref{fig:bio}c.

At 1~kb and $k=12$, 11/12 windows were reconstructed exactly, with 99.9917\% Accuracy and 1.14~GiB RSS. At 2~kb, both $k=14$ and $k=15$ recovered 9/12 windows exactly. Accuracy increased from 99.4917\% to 99.7458\%, while RSS increased from 18.02 to 72.02~GiB. Recorded wall times were 6.92 and 31.55~s, respectively. At 4~kb and $k=15$, 7/12 windows were exact, with 89.5292\% Accuracy and 72.02~GiB RSS. At 8 and 16~kb, none were exact at the tested $k=9$--12 settings. At $k=12$, Accuracy was 53.9250\% and 52.1938\%, respectively. Higher $k$ values at these lengths were not evaluated.

This dense-storage pilot and the expanded 256/128-window comparison used separate input sets. The prediction-preserving hybrid storage tests are reported in Note~\ref{sec:scaling}.

\section{Additional benchmark and scoring experiments}\label{sec:historical-records}

This note reports benchmark quality, timing, the full worker scan and additional scoring experiments. The original benchmark and current A800 comparison were separate batches.

R denotes ACOR runner wall time, including task loading and prediction writing but excluding external evaluation. N denotes historical baseline method-native reconstruction time. A denotes A800 baseline durations summed over clusters or allocated batches, recorded as \src{sum_cluster_seconds}. Because parallel shards can overlap, A is aggregate duration rather than wall time. These records use their stated adapter, preparation and resource settings.
Legacy ACOR times are three-seed medians from the 240-observation worker scan. Legacy baseline times are three-run medians, except for MUSCLE's single diagnostic run. Each A800 cell has one observation. CPL quality was scored from saved batch predictions.
Note~\ref{sec:public-quality} gives the current A800 matrix and batch-specific scoring procedures. Exact and Accuracy are fractions; Mean ED is an average edit count.

\subsection{Historical quality and timing records}
Identical worker-quality columns are shown once. Each duration is identified by dataset, batch, method and timing scope.

\begingroup
\fontsize{10}{12}\selectfont
\setlength{\tabcolsep}{4pt}
\renewcommand{\arraystretch}{1.00}
\setlength{\LTcapwidth}{\linewidth}
\begin{longtable}{@{}llrrr@{}}
\caption{Historical LEGACY-116 quality summaries. Identical ACOR worker-quality columns are shown once. Values are fractions or mean edit counts. Main Fig.~3c uses the separate A800 matrix.}\label{tab:historical-quality}\\
\toprule
Dataset & Method & Exact & Accuracy & Mean ED\\
\midrule
\endfirsthead
\multicolumn{5}{l}{\small\textbf{Table~\thetable\ (continued)}}\\
\toprule
Dataset & Method & Exact & Accuracy & Mean ED\\
\midrule
\endhead
\midrule
\multicolumn{5}{r}{\footnotesize Continued on next page}\\
\endfoot
\bottomrule
\endlastfoot
Oligo0 & ACOR & 0.921555 & 0.997373 & 0.283765\\
Oligo0 & BBS & 0.913370 & 0.997379 & 0.283083\\
Oligo0 & VS & 0.000000 & 0.784479 & 23.276262\\
Oligo0 & BMALA & 0.605730 & 0.985404 & 1.576398\\
Oligo0 & CPL & 0.905866 & 0.998048 & 0.210778\\
Oligo0 & MUSCLE & 0.760573 & 0.985713 & 1.542974\\
ncomms19 365 dishes & ACOR & 0.985280 & 0.997622 & 0.356722\\
ncomms19 365 dishes & BBS & 0.995093 & 0.999748 & 0.037782\\
ncomms19 365 dishes & VS & 0.479392 & 0.963376 & 5.493621\\
ncomms19 365 dishes & BMALA & 0.752699 & 0.992692 & 1.096173\\
ncomms19 365 dishes & CPL & 0.992149 & 0.999617 & 0.057409\\
ncomms19 365 dishes & MUSCLE & 0.985280 & 0.999558 & 0.066241\\
Binned nanopore & ACOR & 0.987500 & 0.998264 & 0.243100\\
Binned nanopore & BBS & 0.988000 & 0.998529 & 0.205900\\
Binned nanopore & VS & 0.739800 & 0.987242 & 1.786100\\
Binned nanopore & BMALA & 0.938200 & 0.997504 & 0.349400\\
Binned nanopore & CPL & 0.984900 & 0.998629 & 0.192000\\
Binned nanopore & MUSCLE & 0.962500 & 0.998160 & 0.257600\\
Microsoft CNR & ACOR & 0.943309 & 0.998303 & 0.186699\\
Microsoft CNR & BBS & 0.947716 & 0.998469 & 0.168369\\
Microsoft CNR & VS & 0.086538 & 0.928032 & 7.916466\\
Microsoft CNR & BMALA & 0.780048 & 0.993643 & 0.699219\\
Microsoft CNR & CPL & 0.948918 & 0.998611 & 0.152744\\
Microsoft CNR & MUSCLE & 0.846955 & 0.997642 & 0.259415\\
Trellis BMA & ACOR & 0.972060 & 0.995559 & 0.488539\\
Trellis BMA & BBS & 0.978842 & 0.999529 & 0.051811\\
Trellis BMA & VS & 0.011122 & 0.924194 & 8.338668\\
Trellis BMA & BMALA & 0.996202 & 0.999921 & 0.008680\\
Trellis BMA & CPL & 1.000000 & 1.000000 & 0.000000\\
Trellis BMA & MUSCLE & 0.905737 & 0.999090 & 0.100095\\
DNAFormer nanopore & ACOR & 0.987525 & 0.998602 & 0.195706\\
DNAFormer nanopore & BBS & 0.987543 & 0.998744 & 0.175902\\
DNAFormer nanopore & VS & 0.738979 & 0.987160 & 1.797601\\
DNAFormer nanopore & BMALA & 0.937359 & 0.997736 & 0.317024\\
DNAFormer nanopore & CPL & 0.984524 & 0.998775 & 0.171446\\
DNAFormer nanopore & MUSCLE & 0.962719 & 0.998383 & 0.226440\\
DNAFormer Illumina & ACOR & 0.999154 & 0.999902 & 0.013734\\
DNAFormer Illumina & BBS & 0.999354 & 0.999924 & 0.010678\\
DNAFormer Illumina & VS & 0.999909 & 0.999998 & 0.000309\\
DNAFormer Illumina & BMALA & 0.999927 & 0.999998 & 0.000327\\
DNAFormer Illumina & CPL & 0.999891 & 0.999998 & 0.000337\\
DNAFormer Illumina & MUSCLE & 0.999927 & 0.999998 & 0.000300\\
Zenodo 10943282 & ACOR & 0.958830 & 0.998542 & 0.218738\\
Zenodo 10943282 & BBS & 0.960202 & 0.998751 & 0.187394\\
Zenodo 10943282 & VS & 0.618525 & 0.984178 & 2.373285\\
Zenodo 10943282 & BMALA & 0.922183 & 0.997243 & 0.413543\\
Zenodo 10943282 & CPL & 0.959950 & 0.998617 & 0.207463\\
Zenodo 10943282 & MUSCLE & 0.938717 & 0.998604 & 0.209428\\
\end{longtable}
\endgroup

\begin{table}[H]\centering
\footnotesize
\setlength{\tabcolsep}{4pt}
\renewcommand{\arraystretch}{1.00}
\caption{ACOR runner wall time on LEGACY-116, in seconds. Scope R includes task loading and prediction writing. Values are three-seed medians.}\label{tab:legacy-acor-times}
\begin{tabular}{@{}lrrrrrrrrrr@{}}
\toprule
Dataset & P1 & P2 & P4 & P8 & P12 & P16 & P20 & P24 & P28 & P32\\
\midrule
Oligo0 & 1.885 & 1.063 & 0.690 & 0.478 & 0.440 & 0.433 & 0.400 & 0.426 & 0.458 & 0.501\\
ncomms19 365 dishes & 1.858 & 1.057 & 0.666 & 0.464 & 0.443 & 0.432 & 0.395 & 0.421 & 0.453 & 0.510\\
Binned nanopore & 5.716 & 3.080 & 1.808 & 1.089 & 0.861 & 0.780 & 0.706 & 0.694 & 0.700 & 0.739\\
Microsoft CNR & 5.608 & 3.007 & 1.811 & 1.090 & 0.870 & 0.782 & 0.677 & 0.693 & 0.709 & 0.775\\
Trellis BMA & 3.701 & 1.987 & 1.169 & 0.717 & 0.583 & 0.535 & 0.472 & 0.490 & 0.525 & 0.558\\
DNAFormer nanopore & 63.131 & 33.193 & 19.240 & 10.948 & 7.917 & 6.616 & 5.691 & 5.544 & 5.413 & 5.311\\
DNAFormer Illumina & 31.932 & 17.451 & 10.678 & 6.667 & 5.346 & 4.728 & 4.241 & 4.135 & 4.128 & 4.190\\
Zenodo 10943282 & 63.423 & 33.211 & 19.302 & 10.914 & 7.917 & 6.621 & 5.655 & 5.442 & 5.320 & 5.254\\
\bottomrule
\end{tabular}
\end{table}

\begin{table}[H]\centering
\fontsize{10}{12}\selectfont
\setlength{\tabcolsep}{4pt}
\renewcommand{\arraystretch}{1.00}
\caption{ACOR runner wall time on A800-173, in seconds. Scope R includes task loading and prediction writing. Each setting has one observation.}\label{tab:a800-acor-times}
\begin{tabular}{@{}lrrrrr@{}}
\toprule
Dataset & P1 & P4 & P16 & P20 & P32\\
\midrule
Oligo0 & 2.277 & 0.731 & 0.399 & 0.421 & 0.478\\
ncomms19 365 dishes & 2.260 & 0.621 & 0.342 & 0.375 & 0.475\\
Binned nanopore & 7.201 & 2.020 & 0.771 & 0.754 & 0.751\\
Microsoft CNR & 7.727 & 1.929 & 0.856 & 0.774 & 0.654\\
Trellis BMA & 4.744 & 1.363 & 0.519 & 0.457 & 0.531\\
DNAFormer nanopore & 73.881 & 19.735 & 6.290 & 5.911 & 4.351\\
DNAFormer Illumina & 33.977 & 10.162 & 4.143 & 3.901 & 3.245\\
Zenodo 10943282 & 195.176 & 24.632 & 6.766 & 9.393 & 7.864\\
\bottomrule
\end{tabular}
\end{table}

\begin{table}[H]\centering
\footnotesize
\setlength{\tabcolsep}{4pt}
\renewcommand{\arraystretch}{1.00}
\caption{Baseline durations in seconds. N is historical method-native time; A is \src{sum_cluster_seconds}, summed durations rather than wall time. Method-specific resource allocations are stated below.}\label{tab:baseline-duration-records}
\begin{tabular}{@{}llrrrrr@{}}
\toprule
Dataset & Batch/scope & BBS & VS & BMALA & CPL & MUSCLE\\
\midrule
Oligo0 & Legacy N & 15.213 & 3.087 & 16.281 & 40.110 & 61247.273\\
Oligo0 & A800 A & 13.175 & 3.917 & 27.402 & 855.551 & 70560.849\\
ncomms19 365 dishes & Legacy N & 15.212 & 5.112 & 9.990 & 58.428 & 4379.504\\
ncomms19 365 dishes & A800 A & 12.574 & 6.215 & 17.046 & 1127.354 & 4563.259\\
Binned nanopore & Legacy N & 51.287 & 26.656 & 18.452 & 132.212 & 6561.279\\
Binned nanopore & A800 A & 42.480 & 32.120 & 48.349 & 2885.637 & 6628.128\\
Microsoft CNR & Legacy N & 43.763 & 8.455 & 22.820 & 110.413 & 11747.941\\
Microsoft CNR & A800 A & 36.771 & 19.856 & 35.507 & 2230.026 & 11650.393\\
Trellis BMA & Legacy N & 39.320 & 3.849 & 6.914 & 19.379 & 951.383\\
Trellis BMA & A800 A & 21.750 & 8.173 & 15.410 & 386.995 & 1105.117\\
DNAFormer nanopore & Legacy N & 504.037 & 146.183 & 197.922 & 1989.921 & 71924.144\\
DNAFormer nanopore & A800 A & 602.559 & 346.166 & 161.637 & 9853.507 & 80174.637\\
DNAFormer Illumina & Legacy N & 280.752 & 271.482 & 163.538 & 1870.090 & 4925.656\\
DNAFormer Illumina & A800 A & 270.352 & 356.412 & 154.719 & 18672.449 & 4808.378\\
Zenodo 10943282 & Legacy N & 695.418 & 262.519 & 248.506 & 2225.972 & 77099.014\\
Zenodo 10943282 & A800 A & 418.267 & 108.470 & 238.702 & 10425.623 & 147116.382\\
\bottomrule
\end{tabular}
\end{table}

Legacy CPL used eight shards and MUSCLE sixteen; A800 VS, BMALA and CPL used eight shards and MUSCLE thirty-two. The DNAFormer nanopore A800 CPL duration sums 56 native invocation times from full-input validation, excluding external evaluation, rather than measuring controller wall time. Note~\ref{sec:public-quality} gives the A800 quality matrix.

\clearpage
\subsection{Historical worker scan and component tests}
Earlier tests compared execution routes and GPU counting. Sparse range merging did not improve some small-cluster timings. Note~\ref{sec:scaling} reports high-$k$ sparse storage and complete CPU reconstruction.

\begin{figure}[H]\centering
\includegraphics[width=0.94\textwidth]{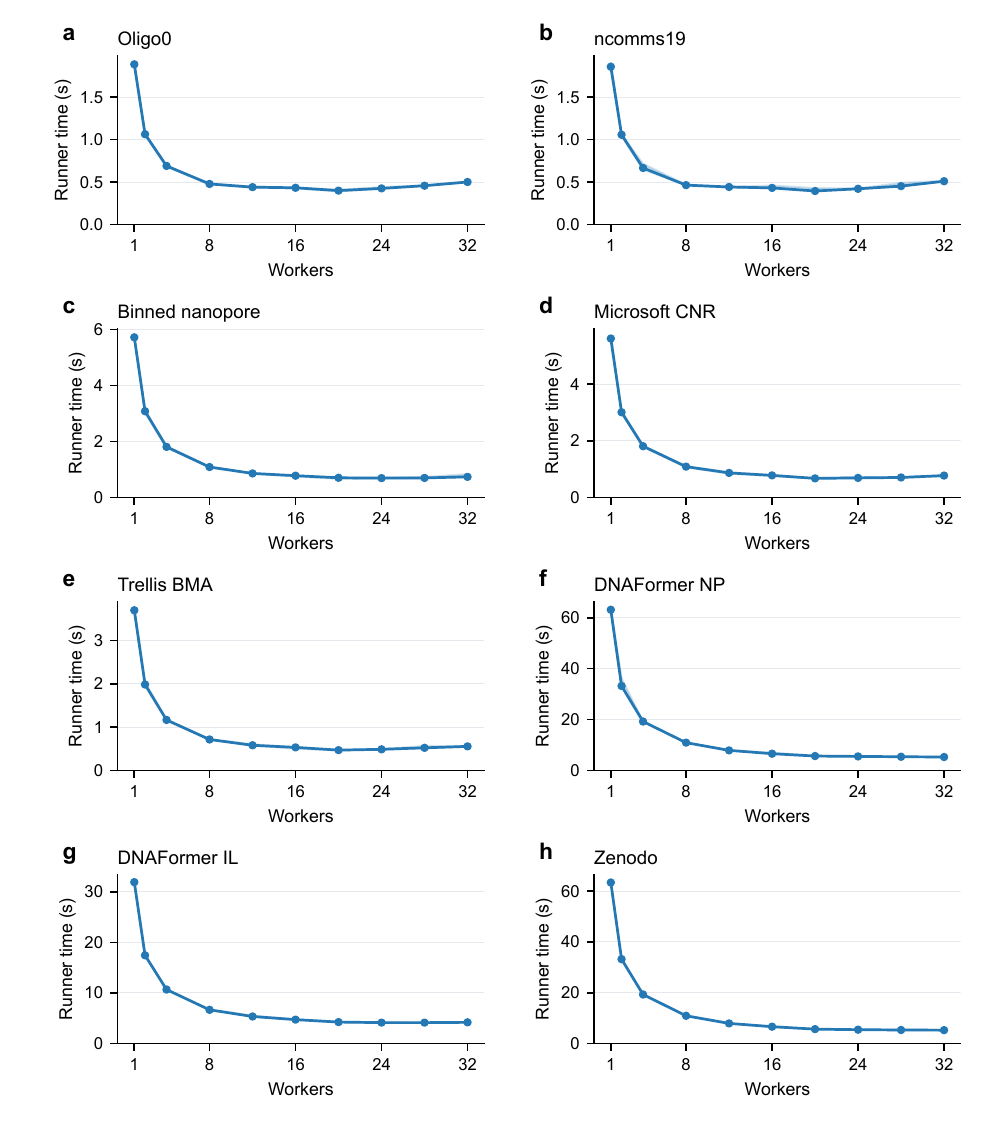}
\caption{Full benchmark worker scan on LEGACY-116. Each panel shows ten worker settings and three recorded seeds for one dataset. Lines are median runner wall times; bands span observed minima--maxima. Runner wall time includes input loading and prediction writing, whereas Fig.~\ref{fig:parallel} also reports reconstruction-compute time separately.}
\label{fig:historical-scaling}
\end{figure}

\subsection{Additional scoring configurations}
\begin{figure}[H]\centering
\includegraphics[width=\textwidth]{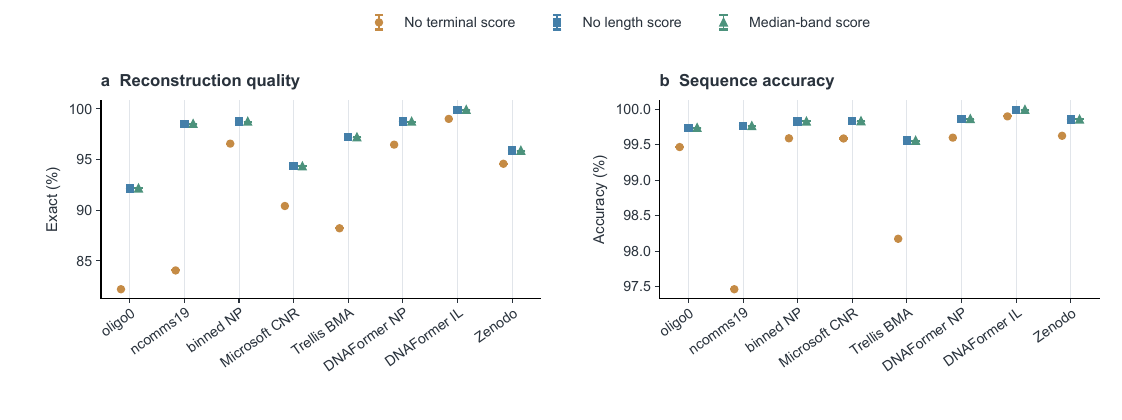}
\caption{Additional scoring variants. \textbf{a}, Exact. \textbf{b}, Accuracy. Points are medians and whiskers are minima and maxima of three recorded repeats for each dataset and variant (72 observations total). The variants remove terminal scoring, remove length scoring, or use median-band scoring. The three variants are compared without a matched default arm.}
\label{fig:score}
\end{figure}

This batch compared three scoring configurations without a matched default arm; main Fig.~3 reports the structural ablations.

\subsection{Parameter-grid source conventions}
For Note~\ref{sec:design-validation}, \src{real_k_grid_full_56.tsv} contains 56 records; \src{synthetic_k_runs.tsv}, 450; \src{synthetic_best_tested_k.tsv}, 90; \src{synthetic_cells.tsv}, 25; and \src{beam_P20.tsv}, 64. \src{score_variants.tsv} contains 72 scoring observations. Aggregated files summarize the corresponding runs. The synthetic \src{error_rate} field denotes $\eta$. Its \src{mean_ed} field stores normalized distance, whereas the main manuscript reports unnormalized Mean ED. Linux RSS values are converted from KiB to GiB by division by $2^{20}$.

Release v0.2.0 exposes fixed $k$ and incorporates the tested engineering optimizations. Measured historical builds are identified in Note~\ref{sec:implementation}, and the release comparison is reported in Note~\ref{sec:release-compatibility}. Notes~\ref{sec:public-quality} and~\ref{sec:own-ngs} give dataset and baseline citations; Note~\ref{sec:design-validation} reports reference-informed selection.

\section{Release compatibility at the default context length}
\label{sec:release-compatibility}
We compared v0.1.0 and v0.2.0 on A800-173 using eight datasets, a fixed read-order seed and default $k=9$. Each version ran once at P1, P20 and P40, giving 48 runs and 24 matched version--dataset--worker pairs. All pairs had identical prediction-file SHA-256 values, Exact, Accuracy and Mean ED.
The source field \texttt{engine\_time\_s} is reported as runner wall time; \texttt{compute\_time\_s} is reported separately. Ratios divide the v0.1.0 observation by the v0.2.0 observation.
At P20, the median runner wall-time ratio across datasets favored v0.2.0 by approximately $1.17\times$. All eight P1 runner observations were slower in v0.2.0, as was Oligo0 at P40. Reconstruction-compute time was also higher in v0.2.0 at the tested worker counts. Timing benefits thus depended on worker count and measurement boundary. Note~\ref{sec:scaling} reports the high-$k$ storage comparison.
\begin{table}[htbp]\centering\fontsize{10}{12}\selectfont
\setlength{\tabcolsep}{3pt}
\renewcommand{\arraystretch}{1.10}
\caption{Default-$k$ version comparison at P1. Each time is one observation; every prediction pair is identical.}
\label{tab:version-p1}
\begin{tabular*}{\textwidth}{@{\extracolsep{\fill}}lrrrrrrr@{}}
\toprule
& \multicolumn{2}{c}{Runner (s)} & \multicolumn{2}{c}{Compute (s)} & \multicolumn{2}{c}{RSS (MiB)} & Runner\\
Dataset & v0.1 & v0.2 & v0.1 & v0.2 & v0.1 & v0.2 & old/new\\
\midrule
Oligo0 & 1.844 & 1.977 & 1.770 & 1.905 & 56.9 & 56.8 & 0.93\\
ncomms19 365 dishes & 1.917 & 1.989 & 1.829 & 1.908 & 44.2 & 44.1 & 0.96\\
Binned nanopore & 5.940 & 6.193 & 5.811 & 6.071 & 74.6 & 74.5 & 0.96\\
Microsoft CNR & 5.852 & 6.114 & 5.712 & 5.975 & 76.0 & 76.2 & 0.96\\
Trellis BMA & 3.954 & 4.083 & 3.888 & 4.020 & 46.2 & 46.4 & 0.97\\
DNAFormer nanopore & 65.127 & 68.162 & 63.747 & 66.795 & 512.4 & 515.3 & 0.96\\
DNAFormer Illumina & 30.558 & 32.684 & 28.840 & 30.983 & 664.0 & 667.3 & 0.93\\
Zenodo 10943282 & 65.453 & 68.110 & 64.103 & 66.778 & 523.0 & 525.7 & 0.96\\
\bottomrule\end{tabular*}
\end{table}
\begin{table}[htbp]\centering\fontsize{10}{12}\selectfont
\setlength{\tabcolsep}{3pt}
\renewcommand{\arraystretch}{1.10}
\caption{Default-$k$ version comparison at P20. Each time is one observation; every prediction pair is identical.}
\label{tab:version-p20}
\begin{tabular*}{\textwidth}{@{\extracolsep{\fill}}lrrrrrrr@{}}
\toprule
& \multicolumn{2}{c}{Runner (s)} & \multicolumn{2}{c}{Compute (s)} & \multicolumn{2}{c}{RSS (MiB)} & Runner\\
Dataset & v0.1 & v0.2 & v0.1 & v0.2 & v0.1 & v0.2 & old/new\\
\midrule
Oligo0 & 0.322 & 0.222 & 0.113 & 0.147 & 410.9 & 408.8 & 1.45\\
ncomms19 365 dishes & 0.332 & 0.239 & 0.102 & 0.139 & 393.2 & 391.6 & 1.39\\
Binned nanopore & 0.579 & 0.493 & 0.314 & 0.362 & 421.8 & 420.8 & 1.17\\
Microsoft CNR & 0.589 & 0.507 & 0.314 & 0.356 & 423.2 & 422.9 & 1.16\\
Trellis BMA & 0.410 & 0.316 & 0.210 & 0.253 & 393.4 & 392.6 & 1.30\\
DNAFormer nanopore & 5.074 & 5.090 & 3.515 & 3.661 & 861.0 & 864.3 & 1.00\\
DNAFormer Illumina & 3.431 & 3.403 & 1.516 & 1.649 & 1010.2 & 1012.7 & 1.01\\
Zenodo 10943282 & 5.078 & 5.039 & 3.534 & 3.650 & 870.7 & 873.6 & 1.01\\
\bottomrule\end{tabular*}
\end{table}
\begin{table}[htbp]\centering\fontsize{10}{12}\selectfont
\setlength{\tabcolsep}{3pt}
\renewcommand{\arraystretch}{1.10}
\caption{Default-$k$ version comparison at P40. Each time is one observation; every prediction pair is identical.}
\label{tab:version-p40}
\begin{tabular*}{\textwidth}{@{\extracolsep{\fill}}lrrrrrrr@{}}
\toprule
& \multicolumn{2}{c}{Runner (s)} & \multicolumn{2}{c}{Compute (s)} & \multicolumn{2}{c}{RSS (MiB)} & Runner\\
Dataset & v0.1 & v0.2 & v0.1 & v0.2 & v0.1 & v0.2 & old/new\\
\midrule
Oligo0 & 0.378 & 0.853 & 0.060 & 0.779 & 782.4 & 778.1 & 0.44\\
ncomms19 365 dishes & 0.388 & 0.221 & 0.054 & 0.136 & 760.3 & 757.5 & 1.76\\
Binned nanopore & 0.529 & 0.367 & 0.160 & 0.252 & 788.3 & 786.9 & 1.44\\
Microsoft CNR & 0.536 & 0.393 & 0.160 & 0.254 & 789.8 & 788.9 & 1.37\\
Trellis BMA & 0.415 & 0.252 & 0.107 & 0.192 & 759.3 & 754.9 & 1.65\\
DNAFormer nanopore & 3.333 & 3.248 & 1.697 & 1.855 & 1226.4 & 1223.2 & 1.03\\
DNAFormer Illumina & 2.738 & 2.622 & 0.762 & 0.901 & 1374.2 & 1373.4 & 1.04\\
Zenodo 10943282 & 3.310 & 3.221 & 1.700 & 1.861 & 1236.5 & 1237.3 & 1.03\\
\bottomrule\end{tabular*}
\end{table}

\begin{figure}[ht!]\centering
\includegraphics[width=\textwidth]{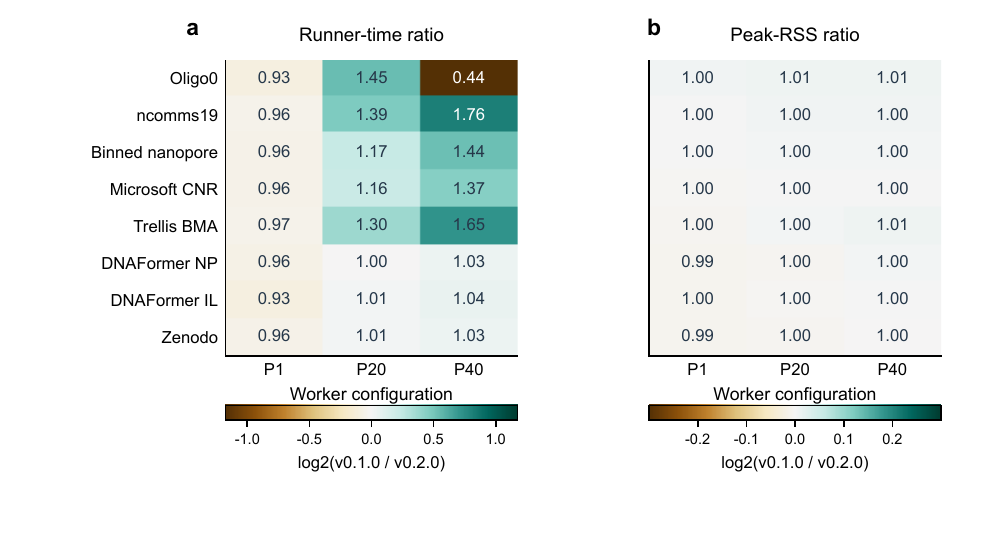}

\caption{Default-$k$ release comparison. \textbf{a}, Runner wall-time ratios. \textbf{b}, Peak-RSS ratios. Values divide v0.1.0 by v0.2.0; colours show their base-2 logarithms. Values above one favor v0.2.0 for the displayed quantity. Each of the 24 pairs contains one observation per version. All paired predictions were identical. Source Data: \src{v010_v020_a800.tsv}.}
\label{fig:release}
\end{figure}

\clearpage
\section*{Source Data index}\label{sec:source-data-index}
\addcontentsline{toc}{section}{Source Data index}

Experimental records are supplied in \src{ACOR_Source_Data.zip}. Archive paths and checksums are listed in \src{_archive_metadata/FILES.tsv}. Navigation maps are \src{SourceData/main_figure_table_map.tsv} and \src{SourceData/supplement_figure_table_map.tsv}.

\textbf{Parameter and component comparisons.} \src{synthetic_k_runs.tsv} contains 450 runs; \src{real_k_grid_full_56.tsv}, 56 configurations; \src{beam_P20.tsv}, 64 observations; and \src{score_variants.tsv}, 72 scoring-variant observations. The reference-informed assessment uses \src{direct_k_selector_validation.tsv}, \src{real_dataset_profiles.tsv} and \src{real_truth_kmer_structure.tsv}. Paired support-weight and structural-ablation records remain separate from these single-configuration grids.

\textbf{Public benchmark quality and timing.} The A800 matrix has 48 cells and an 80-record table; the original worker scan has 240 observations. Records identify each batch and measured implementation. The full-input CPL run is in \src{SourceData/cpl_verified_20261013/}.

\textbf{Own NGS library.} Assigned reads, within-cluster order, design references and coverage are in \src{NGS/}. JSON records for 22 configurations and four historical method snapshots are mapped in \src{SourceData/provenance/ngs_native_record_index.tsv}. References were used for assignment and evaluation, not reconstruction. The assigned-read archive excludes raw paired FASTQ files.

\textbf{Storage and CPU scaling.} \src{HYBRID_MATRIX.tsv} and \src{high_k_sparse_matrix.tsv} contain the same 32-configuration matrix. \src{parallel_5m_runs.tsv} and \src{parallel_20m_runs.tsv} contain 27 and 36 timing observations; their \src{engine_seconds} field records reconstruction-compute time. The 72 benchmark output-identity checks are a separate series.

\textbf{Long-sequence experiments.} \src{REAL_ONT_LATEST.tsv} contains nine ACOR length--worker configurations. \src{REAL_ONT_LATEST_COMPARISON.tsv} reports 14 completed method--length cells and one recorded timeout. The \src{ont/} records retain input construction, repeated baseline measurements, error distributions and stopping states. The historical pilot is stored separately in \src{real_bio_k_grid.tsv} and \src{real_bio_window_manifest.tsv}.

\textbf{Release compatibility.} \src{v010_v020_a800.tsv} contains 48 version-comparison runs; \src{v010_v020_paired_summary_24.tsv} summarizes them as 24 matched pairs.

Measured-build and executable identifiers are listed in \src{SourceData/provenance/experiment_build_identities.tsv}.

\end{document}